\documentclass[smallextended]{svjour3}
\usepackage{graphicx, amsmath, amssymb, hyperref, tikz, booktabs, multirow}
\usepackage[caption=false]{subfig}

\usepackage[leftcaption]{sidecap}
\sidecaptionvpos{figure}{t}

\usepackage{enumitem}
\setlist[enumerate]{parsep=0.5em,itemsep=0em,topsep=0.5em}

\allowdisplaybreaks[1]

\newcommand{\braket}[2]{{\left\langle #1 \middle| #2 \right\rangle}}

\newcommand{\ket}[1]{{\left| #1 \right\rangle}}
\newcommand{\ketbra}[2]{{\left| #1 \middle\rangle \middle \langle #2 \right|}}
\newcommand{\fref}[1]{Fig.~\ref{#1}}

\newcommand{\sref}[1]{Section~\ref{#1}}
\newcommand{\tref}[1]{Table~\ref{#1}}

\journalname{Quantum Inf Process}

\begin{document}

\title{Optimal and Deterministic Quantum Search on the Simplex of Complete Graphs}

\author{Kiyoji Huang Fujiwara$^1$ \and Yujia Shi$^1$ \and Thomas G.~Wong$^1$}

\authorrunning{K.~Huang Fujiwara \and Y.~Shi \and T.~G.~Wong}

\institute{$^1$Department of Physics, Creighton University, 2500 California Plaza, Omaha, Nebraska 68178, USA \\
	\email{kiyojihuangfujiwara@creighton.edu} \\
    \email{yujiashi@creighton.edu} \\
	\email{thomaswong@creighton.edu}
}

\date{Received: date / Accepted: date}

\maketitle

\begin{abstract}
	The simplex of complete graphs, also known as the first-order truncated simplex lattice, is a network of $M+1$ identical complete graphs, each with $M$ vertices, such that each clique contains an edge or bridge to every other clique. It contains $N = M(M+1)$ vertices, and previous asymptotic results using a continuous-time quantum walk to search this graph for a single marked vertex have either numerically demonstrated an optimal runtime of $O(\sqrt{N})$, or analytically proved a deterministic success probability of 1, but not both, even when the bridges are weighted. In this paper, we give the first analytical proof of optimal quantum search on this graph, proving that it occurs when the weight of the bridges equals $M$. In addition, we numerically show that the optimal runtime is achieved more broadly whenever the weight is at least $\sqrt{M}$. Furthermore, the algorithm is also deterministic when the weight scales between $\sqrt{M}$ and $M$, and this is the first example of quantum search on the simplex of complete graphs that is both asymptotically optimal and deterministic. In addition, for weights where the algorithm is nondeterministic, we give a way to find the marked vertex by inspecting neighboring vertices. Finally, while it is known that connectivity is not a reliable indicator of fast quantum search when comparing different graph families, we show that it is also unreliable within the graph family of weighted simplex of complete graphs.
\end{abstract}


\section{Introduction}

One of the first algorithms developed for quantum computers was Grover's quantum search algorithm \cite{Grover1996}, which alternates between querying an oracle and applying quantum gates in order to search an unordered database of $N$ items in $O(\sqrt{N})$ time, which is optimal \cite{Boyer1998}. However, for data that is physically distributed, it takes time to move through the database, and the optimal runtime of $O(\sqrt{N})$ might not be achievable \cite{Benioff2002}. To solve this spatial search problem, many solutions have been proposed \cite{AA2005}, but one of the most common involves quantum walks \cite{CG2004,SKW2003}, which are quantum analogs of random walks \cite{Kempe2003}.

In a quantum walk, the $N$ vertices of a graph encode the $N$ items of the database, and edges between vertices indicate allowed movements from one item to another. Labeling the $N$ vertices as computational basis states $\ket{1}, \ket{2}, \dots, \ket{N}$, the state $\ket{\psi(t)}$ of the search algorithm begins in a uniform superposition over these basis states, which we call $\ket{s}$, i.e.,
\begin{equation}
    \label{eq:s}
    \ket{\psi(0)} = \ket{s} = \frac{1}{\sqrt{N}} \sum_{i=1}^{N} \ket{i}.
\end{equation}
This equal superposition expresses our initial lack of knowledge about the marked vertex's location by assigning an equal probability to each vertex. In a continuous-time quantum walk \cite{FG1998a}, the state evolves by Schr\"odinger's equation,
\[ i \hbar \frac{d}{dt} \ket{\psi(t)} = H \ket{\psi}, \]
where we take the reduced Planck constant $\hbar = 1$ throughout this paper, and $H$ is the Hamiltonian. For the search algorithm \cite{CG2004}, the Hamiltonian consists of two terms, one that drives the quantum walk \cite{FG1998a}, and another that serves as the oracle marking the vertex to be found \cite{Mochon2007}:
\begin{equation}
    \label{eq:H}
    H = -\gamma A - \ketbra{a}{a},
\end{equation}
where $\gamma$ is the jumping rate of the quantum walk, $A$ is the adjacency matrix of the graph (where the matrix entry $A_{ij}$ is the weight of the edge joining vertices $i$ and $j$ if they are adjacent, or 0 if they are nonadjacent), and $\ket{a} \in \{ \ket{1}, \dots, \ket{N} \}$ is the marked vertex we are looking for. Since the search Hamiltonian \eqref{eq:H} is time independent, the solution to Schr\"odinger's equation (with $\hbar = 1$) is
\begin{equation}
    \label{eq:psi-t}
    \ket{\psi(t)} = e^{-iHt} \ket{\psi(0)}.
\end{equation}
This quantum walk-based search algorithm was introduced in \cite{CG2004} for searching the complete graph, hypercube, and arbitrary-dimensional periodic square lattices, and it has since been explored for various graphs, some of which include strongly regular graphs \cite{Wong5}, complete bipartite graphs \cite{Novo2015,Wong19}, trees \cite{Philipp2016}, Johnson graphs \cite{Wong20,Tanaka2022}, and infinite graphs \cite{Xie2023}.

\begin{SCfigure}
    \caption{\label{fig:simplex}A simplex of complete graphs, where each clique has $M = 5$ vertices, so there are $N = 30$ vertices in total. The solid edges within each clique have weight 1, and the dashed bridges between cliques have weight $w$. A vertex is marked, as indicated by a double circle. Due to the symmetry of the graph, some vertices evolve with the same amplitude, and identically evolving vertices are identically colored and labeled.}
    \includegraphics{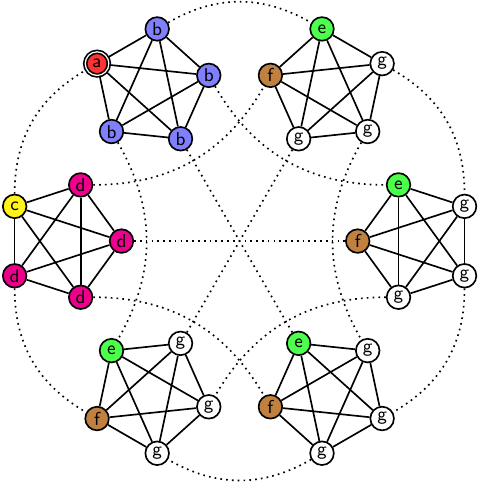}
\end{SCfigure}

In this paper, we focus on the simplex of complete graphs \cite{Wong7}, also called the first-order truncated simplex lattice \cite{Dhar1977}. An example is shown in \fref{fig:simplex}. It consists of $M+1$ complete graphs, each with $M$ vertices, so the total number of vertices is $N = M(M+1)$. In \fref{fig:simplex}, $M = 5$. In each clique, every vertex is adjacent to a unique vertex in a unique other clique, as shown in \fref{fig:simplex}. Furthermore, we permit the graph to be weighted such that the inter-cluster edges or bridges each have weight $w \ge 1$, corresponding to the dotted edges in \fref{fig:simplex}, while the intra-cluster edges remain unweighted (i.e., have weight 1). This way, the graph is vertex transitive, meaning each vertex has the same structure, so no matter which vertex is marked, the system will evolve in the same manner to the marked vertex.

\begin{figure}
\begin{center}
    \subfloat[] {
        \includegraphics{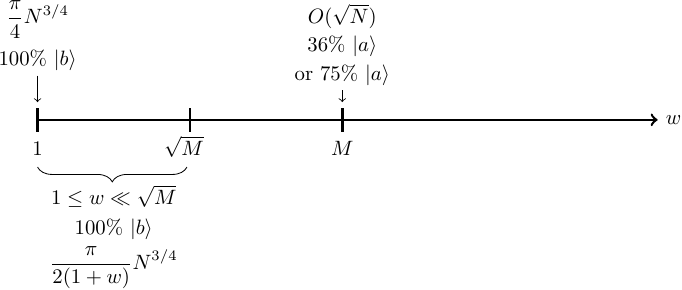}
        \label{fig:summary-priorwork}
    }
    
    \subfloat[] {
        \includegraphics{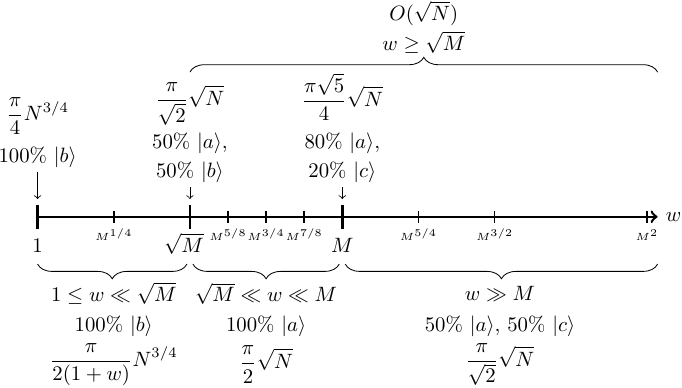}
        \label{fig:summary}
    }
    \caption{Summary the first (or only) stage for search on the weighted simplex of complete graphs for large $N$. Subfigure (a) depicts only the prior results, and (b) includes the present results. The minor ticks in (b) are examples that are investigated in the paper.}
\end{center}
\end{figure}

In the context of spatial search, the simplex of complete graphs was first introduced with $w = 1$ to disprove the intuition that high connectivity results in faster quantum search \cite{Wong7}. To prove this, it was shown that a two-stage algorithm was needed, with each stage chosen by selecting the jumping rate $\gamma$ that appears in the Hamiltonian \eqref{eq:H}. The first stage takes $\gamma = 2/M$ and evolves the system from the initial uniform state $\ket{s}$ \eqref{eq:s} to a uniform state over the $b$ vertices shown in \fref{fig:simplex}, which are in same clique as the marked vertex, and this process takes $\pi N^{3/4}/4$ time, for large $N$. This first stage is summarized in \fref{fig:summary-priorwork}, which depicts the results for various $w$. It shows a ray of possible values of $w$, and the leftmost tick corresponds to $w = 1$. Above this endpoint is the aforementioned asymptotic first-stage runtime of $\pi N^{3/4}/4$ and resulting state of $\ket{b}$, where $\ket{b}$ is a uniform superposition over the $b$ vertices. This first stage is followed by a second stage that completes the search. In the second stage, the jumping rate is changed to $\gamma = 1/M$, causing the walker to further focus from $\ket{b}$ to the marked vertex $\ket{a}$, taking an additional $\pi N^{1/4}/2$ time, for large $N$. Then, measuring the position of the walker, it will be found at the marked vertex $\ket{a}$ with asymptotic probability 1, so the algorithm is deterministic, for large $N$. Together, the overall runtime is dominated by the first stage's runtime of $\pi N^{3/4}/4 = O(N^{3/4})$, which is relatively slow for the high connectivity of the graph.

In \cite{Wong16}, the weighted ($w > 1$) simplex of complete graphs was introduced. It was shown that when $w$ scales less than $\sqrt{M}$, i.e., $w \ll \sqrt{M}$ or $w = o(\sqrt{M})$ in asymptotic notation, the two-stage algorithm remains. In the first stage, a jumping rate of $\gamma = (1+1/w)/M$ causes the system to evolve from $\ket{s}$ to $\ket{b}$ in time $\pi N^{3/4}/[2(1+w)]$, and so the greater the weight $w$, the faster the speedup. This is summarized in \fref{fig:summary-priorwork}, where the region where $1 \le w \ll \sqrt{M}$ is underbraced and labeled with the asymptotic resulting state of $\ket{b}$ and runtime of $\pi N^{3/4}/[2(1+w)]$. Since this result requires that $w$ scale less than $\sqrt{M}$, this first-stage runtime can almost drop as low as $O(\sqrt{N})$, but not quite. The second stage remains the same as the unweighted case. That is, using a jumping rate of $\gamma = 1/M$, the system evolves from $\ket{b}$ to $\ket{a}$ in time $\pi N^{1/4}/2$, so the two-stage algorithm is deterministic, asymptotically. The overall runtime is dominated by this first stage, and so the overall runtime of $O(N^{3/4}/w)$ is short of being optimal since $w \ll \sqrt{M}$.

With stronger weights that scale as or beyond $\sqrt{M}$, \cite{Wong16} argued that the two stages would be merged into a single stage, as the jumping rates of the two stages would collide. Investigating the behavior of such single-stage algorithms, however, was left for further research. Such further work was conducted by Wang, Wu, and Wang \cite{Wang2020}, who focused on the algorithm's behavior when $w$ scales greater than $M^{3/4}$, i.e., $w \gg M^{3/4}$ or $w = \omega(M^{3/4})$ in asymptotic notation. Using degenerate perturbation theory, they proposed the following jumping rate for the single-stage search algorithm when $w \gg M^{3/4}$:
\begin{equation}
    \label{eq:gamma-WWW}
    \gamma_\text{\tiny WWW} = \frac{M+w}{M(M+2w)}.
\end{equation}
They only explored this for the special case of $w = M$, for which $\gamma_\text{\tiny WWW} = 2/(3M)$. They numerically showed that the probability of measuring the walker at the marked vertex $\ket{a}$, i.e., the success probability, reaches 36\% in optimal $O(\sqrt{N})$ time. Numerically correcting the jumping rate to $\gamma = 2/(3M) + 2/M^2$, however, resulted in a higher success probability of 75\% in optimal $O(\sqrt{N})$ time. These results are summarized in \fref{fig:summary-priorwork}, where listed above the $w = M$ tick mark is the $O(\sqrt{N})$ runtime and the success probability of 36\% or 75\%, depending on which jumping rate is used. Although these algorithms have optimal runtimes, they are nondeterministic.

\begin{figure}
\begin{center}
    \subfloat[Probability in $\ket{a}$.] {
        \includegraphics[height=1.85in]{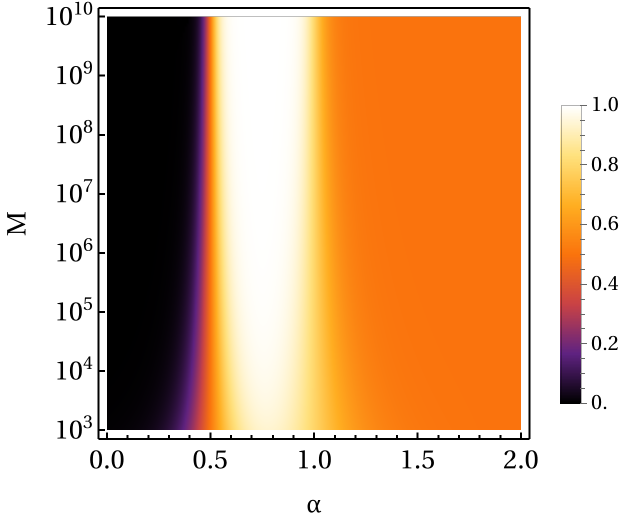}
        \label{fig:heatmap-a}
    }
    \subfloat[Probability in $\ket{b}$.] {
        \includegraphics[height=1.85in]{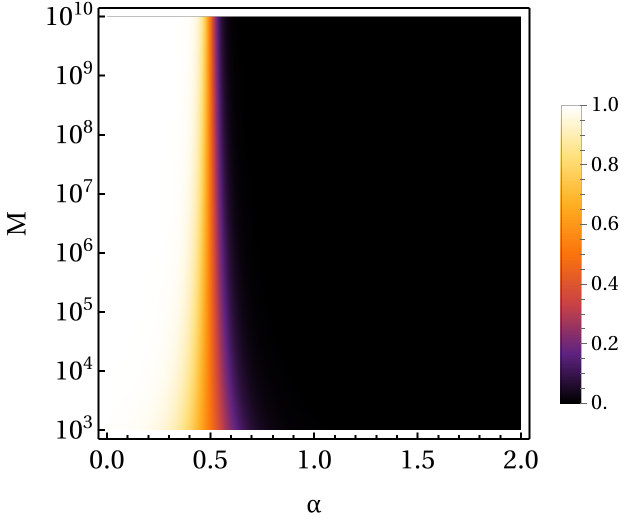}
        \label{fig:heatmap-b}
    }
    
    \subfloat[Probability in $\ket{c}$.] {
        \includegraphics[height=1.85in]{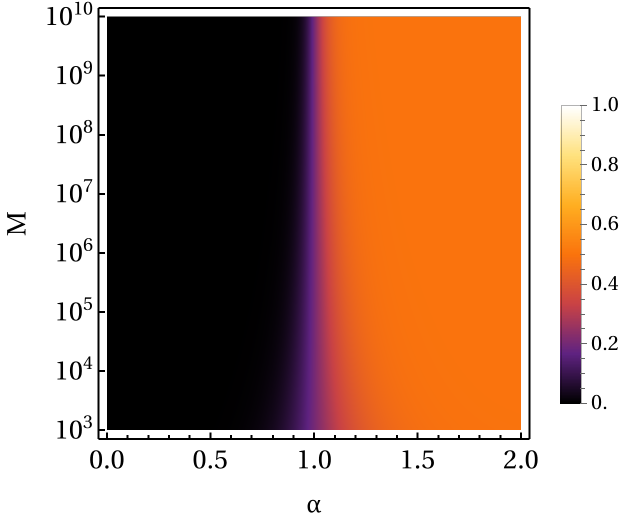}
        \label{fig:heatmap-c}
    }
    \subfloat[$\text{Runtime}/\sqrt{N}$.] {
        \includegraphics[height=1.85in]{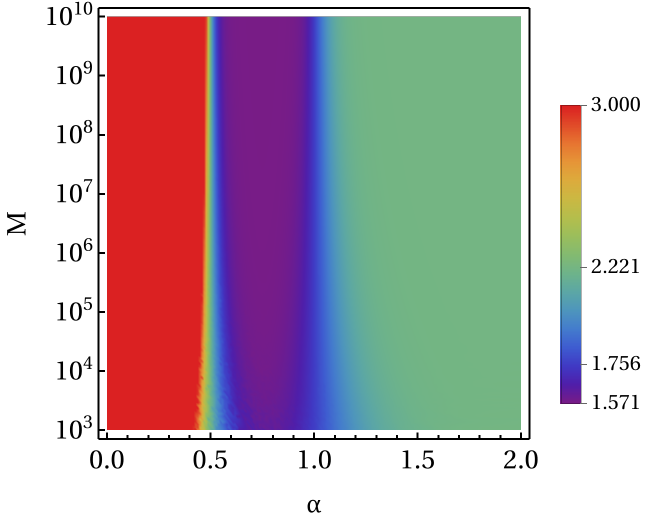}
        \label{fig:heatmap-runtime}
    }
    \caption{\label{fig:heatmaps}For search on the simplex of complete graph with $M(M+1)$ vertices (ranging from $M = 10^3$ to $10^{10}$) and $w = M^\alpha$ (ranging from $w = M^0$ to $M^2$), the probability reached in (a) $\ket{a}$, (b) $\ket{b}$, and (c) $\ket{c}$ at the numerically-determined runtime using the numerically determined critical jumping rate. Subfigure (d) is the numerically-determined runtime divided by $\sqrt{N}$, and the color red is representative of values $\geq 3.000$.}
\end{center}
\end{figure}

It is also possible to consider search on the simplex of complete graphs with multiple marked vertices \cite{Wong9,Wong11,Zhu2023}, or search on higher-order truncated simplex lattices \cite{Wang2020,Zhu2023,Zhang2024}. In this paper, however, we focus on the simplex of complete graphs, i.e., the first-order truncated simplex lattice, with a single marked vertex. In this setting, \fref{fig:summary-priorwork} reveals that most weights $w$ were unexplored in prior work. In this paper, we address this by exploring search on a wide range of weights $w = M^\alpha$ (where $\alpha \ge 0$) as $M$ gets progressively larger. The resulting probabilities in $\ket{a}$, $\ket{b}$, and $\ket{c}$ (c.f., \fref{fig:simplex}) are respectively shown in \fref{fig:heatmap-a} through \fref{fig:heatmap-c}, while the runtime divided by $\sqrt{N}$ is shown in \fref{fig:heatmap-runtime}. From these, we draw several conclusions that are summarized in \fref{fig:summary}, namely:
\begin{enumerate}
    \item When $w = M$, then $\alpha = 1$, and \fref{fig:heatmap-a} shows that the state reaches 80\% $\ket{a}$, and \fref{fig:heatmap-c} shows that the remaining 20\% is in $\ket{c}$. In \sref{sec:w=M}, we will prove this analytically using degenerate perturbation theory with a change of basis, including deriving the critical jumping rate that accomplishes this, as well as the runtime of $\pi\sqrt{5N}/4 \approx 1.756 \sqrt{N}$, which agrees with \fref{fig:heatmap-runtime}. These results are shown above the $w = M$ tick mark in \fref{fig:summary}. This is the first analytical proof of optimal quantum search on the simplex of complete graphs, as the previous results of \cite{Wang2020} were numerical.
    
    The remaining results below are numerical.

    \item When $w = \sqrt{M}$, \fref{fig:heatmap-a} and \fref{fig:heatmap-b} with $\alpha = 0.5$ indicate that the quantum walk asymptotically evolves to 50\% $\ket{a}$ and 50\% $\ket{b}$, and \fref{fig:heatmap-runtime} suggests that the runtime is $\pi\sqrt{N}/\sqrt{2} \approx 2.221 \sqrt{N}$, as noted in \fref{fig:summary} above the tick mark for $w = \sqrt{M}$. This is explored more systematically in \sref{sec:w=M0.5}.

    \item When $w$ is between 1 and $\sqrt{M}$, \fref{fig:heatmap-a} and \fref{fig:heatmap-b} with $0 \le \alpha < 0.5$ show that as $M$ increases, the proportion in $\ket{a}$ decreases, while the proportion in $\ket{b}$ increases. So, within this region, the state asymptotically evolves to just $\ket{b}$. This is the previous result from \cite{Wong16}, which appears in both \fref{fig:summary-priorwork} and \fref{fig:summary} by the underbrace with ``$100\%\ \ket{b}$'' in both. We will also illustrate this in \sref{sec:w<<M^0.5} using $w = M^{1/4}$ as an example, and in \fref{fig:summary}, the minor ticks denote examples given in this paper.

    \item For $\sqrt{M} \ll w \ll M$, as $M$ gets larger, \fref{fig:heatmap-a} through \fref{fig:heatmap-c} show that the system evolves solely to the marked vertex $\ket{a}$, and \fref{fig:heatmap-runtime} indicates that the runtime is $\pi\sqrt{N}/2 \approx 1.571\sqrt{N}$. This is indicated in \fref{fig:summary} by the underbrace with ``$100\%\ \ket{a}$'' and the runtime. Several examples will be given in \sref{sec:M^0.5<<w<<M}, indicated by the minor ticks in \fref{fig:summary}. Thus, this whole region yields search algorithms that are asymptotically optimal and deterministic.

    \item When $w \gg M$, as $M$ increases, \fref{fig:heatmap-a} and \fref{fig:heatmap-c} indicate that when the system evolves to half $\ket{a}$ and half $\ket{c}$. Furthermore, \fref{fig:heatmap-runtime} shows that the runtime is $\pi\sqrt{N/2} \approx 2.221 \sqrt{N}$. These results are summarized in \fref{fig:summary} as an underbrace below the region where $w \gg M$. Examples will be given \sref{sec:w>>M}, indicated by minor ticks in \fref{fig:summary}.
    
    \item When $w \ge \sqrt{M}$, \fref{fig:heatmap-runtime} shows that runtime is asymptotically $O(\sqrt{N})$, although the constant factor may be different depending on $w$. This is indicated in \fref{fig:summary} by the topmost overbrace with $O(\sqrt{N})$ above it.

    In this regime, note that the success probability is always at least 50\%. Then, we expect that we will need to repeat the algorithm at most twice before finding the marked vertex, on average. Then, the expected overall runtime is still $O(\sqrt{N})$, so we have an optimal search algorithm whenever $w \ge \sqrt{M}$. This greatly expands the known weights for which search is optimal. That is, from \fref{fig:summary-priorwork}, an optimal algorithm was only previously known for $w = M$, and now from \fref{fig:summary}, the algorithm is optimal whenever $w$ is at least $\sqrt{M}$.
\end{enumerate}
Typically, the weight $w$ is considered a tunable parameter \cite{Wong22,Wong41}, and so the best choice is $\sqrt{M} \ll w \ll M$, such as $w = M^{3/4}$, for which the algorithm has an optimal runtime with the smallest constant-factor of $\pi/2$, and is deterministic. That there exists an optimal weight for the bridges has been observed for other graphs, namely the Cartesian product of the complete graph with a weighted path of length 2 \cite{Wong22}, and the weighted barbell graph \cite{Wong41,Wong45}.

As discussed above, when the success probability is less than 1, it may be necessary to repeat the algorithm. In \sref{sec:neighbors}, we will give an alternative method to finding the marked vertex that does not require repeating the algorithm. Instead, it involves inspecting neighboring vertices of the measured one.

Since the simplex of complete graphs was introduced in quantum computing to explore connectivity \cite{Wong7}, in \sref{sec:connectivity}, we revisit this topic in the context of our broader results. We will see that the performance of the search algorithm does not necessarily correlate with the connectivity of the graph.

Let us now proceed with \sref{sec:subspace}, where we will review results from \cite{Wong7,Wong16} formulating the quantum walk in a seven-dimensional (7D) subspace. This will lay the groundwork for \sref{sec:w=M} through \sref{sec:w>>M}, where we explore various aspects of \fref{fig:summary} that we enumerated above. Then, we propose a method for finding the marked vertex when the success probability is less than 1 in \sref{sec:neighbors} and discuss connectivity in \sref{sec:connectivity}. Finally, we conclude in \sref{sec:conclusion}.


\section{\label{sec:subspace}Subspace Evolution}

As noted in \cite{Wong7,Wong16}, from the symmetry of the problem, the amplitude at many of the vertices are identical. In fact, there are only seven different amplitudes, corresponding to seven types of vertices, depicted in \fref{fig:simplex}. For example, all the blue $b$ vertices evolve in the same way, so they can be represented by a single basis state. Thus, we can group identically evolving vertices together to form a basis for a 7D subspace:
\begin{align*}
    \ket{a} &= \ket{\rm red}, \\
    \ket{b} &= \frac{1}{\sqrt{M-1}} \sum_{i \in {\rm blue}} \ket{i}, \\
    \ket{c} &= \ket{\rm yellow}, \\
    \ket{d} &= \frac{1}{\sqrt{M-1}} \sum_{i \in {\rm magenta}} \ket{i}, \\
    \ket{e} &= \frac{1}{\sqrt{M-1}} \sum_{i \in {\rm green}} \ket{i}, \\
    \ket{f} &= \frac{1}{\sqrt{M-1}} \sum_{i \in {\rm brown}} \ket{i}, \\
    \ket{g} &= \frac{1}{\sqrt{(M-1)(M-2)}} \sum_{i \in {\rm white}} \ket{i}.
\end{align*}
Note that for a large number of vertices, the initial state $\ket{s} \approx \ket{g}$, since the $g$ vertices vastly outnumber the other types of vertices.

From \cite{Wong7}, in this 7D subspace, the initial equal superposition state \eqref{eq:s} is
\begin{align}
    \ket{s} 
        = \frac{1}{\sqrt{N}} \Big( &\ket{a} + \sqrt{M-1}\ket{b} + \ket{c} + \sqrt{M-1}\ket{d} + \sqrt{M-1}\ket{e} \nonumber \\
        &+ \sqrt{M-1}\ket{f} + \sqrt{(M-1)(M-2)}\ket{g} \Big), \label{eq:s-7D}
\end{align}
and from \cite{Wong16}, the search Hamiltonian is
\begin{equation}
    \label{eq:H-7D}
    H = -\gamma \begin{pmatrix}
        \frac{1}{\gamma} & \sqrt{M_1} & w & 0 & 0 & 0 & 0 \\
        \sqrt{M_1} & M_2 & 0 & 0 & w & 0 & 0 \\
        w & 0 & 0 & \sqrt{M_1} & 0 & 0 & 0 \\
        0 & 0 & \sqrt{M_1} & M_2 & 0 & w & 0 \\
        0 & w & 0 & 0 & 0 & 1 & \sqrt{M_2} \\
        0 & 0 & 0 & w & 1 & 0 & \sqrt{M_2} \\
        0 & 0 & 0 & 0 & \sqrt{M_2} & \sqrt{M_2} & M_3+w
    \end{pmatrix},
\end{equation}
where $M_i = M - i$. Thus, we have reduced an $N$-dimensional problem to a 7D one, where the evolution \eqref{eq:psi-t} can be found using the above 7D initial state \eqref{eq:s-7D} and Hamiltonian \eqref{eq:H-7D}. In the next three sections, we consider the search algorithm in this 7D subspace with with various weight bridges.


\section{\label{sec:w=M}Search when \texorpdfstring{$w = M$}{w=M}}
 
In this section, we explore search on the simplex of complete graphs when the bridges have weight $w = M$.

\begin{figure}
\begin{center}
    \subfloat[] {
        \includegraphics[width=2.25in]{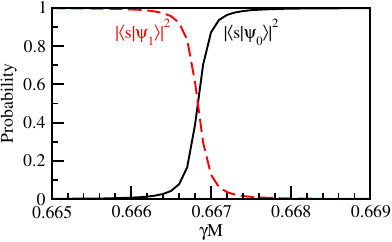}
        \label{fig:overlaps-w=M-s}
    }
    \subfloat[] {
        \includegraphics[width=2.25in]{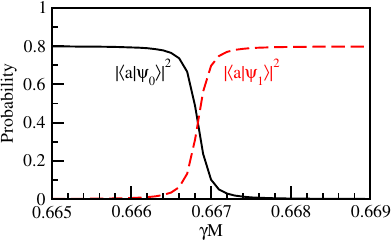}
        \label{fig:overlaps-w=M-a}
    }

    \subfloat[] {
        \includegraphics[width=2.25in]{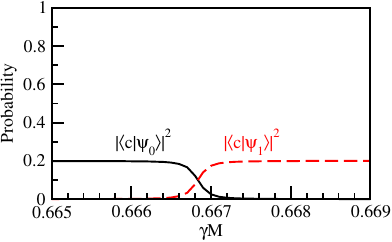}
        \label{fig:overlaps-w=M-c}
    }
    \caption{\label{fig:overlaps-w=M}For the simplex of complete graphs with $M = 10\,000$ and $w = M$, probability overlaps of (a) the initial state $\ket{s}$, (b) the marked vertex $\ket{a}$, and (c) $\ket{c}$ with two eigenvectors $\ket{\psi_0}$ (solid black) and $\ket{\psi_1}$ (dashed red) of the search Hamiltonian for various values of $\gamma M$.}
\end{center}
\end{figure}

First, let us show that the behavior of the algorithm depends on the jumping rate $\gamma$ \eqref{eq:H}. Following \cite{CG2004}, we calculate the eigenvectors of the search Hamiltonian \eqref{eq:H-7D}, which we label $\ket{\psi_0}, \ket{\psi_1}, \dots, \ket{\psi_6}$ for various values of $\gamma$. Then, in \fref{fig:overlaps-w=M}, we calculate the norm-square of inner product of two eigenvectors of interest, $\ket{\psi_0}$ and $\ket{\psi_1}$, with the initial state $\ket{s}$ in \fref{fig:overlaps-w=M-s}, the marked vertex $\ket{a}$ in \fref{fig:overlaps-w=M-a}, and the $\ket{c}$ vertex in \fref{fig:overlaps-w=M-c}. We see from the left side of \fref{fig:overlaps-w=M-s} that when $\gamma$ is small, the initial state $\ket{s}$ is approximately equal to the eigenvector $\ket{\psi_1}$, and so the system does not evolve apart from a global phase. Similarly, on the right side, when $\gamma$ is large, the initial state $\ket{s}$ is approximately equal to the eigenvector $\ket{\psi_0}$, and again the system does not evolve apart from a global phase. Thus, for the system to evolve away from the initial state $\ket{s}$, the jumping rate $\gamma$ must take an intermediate ``critical'' value around the middle of the plot, where the curves cross around $\gamma M = 0.6668$. From \fref{fig:overlaps-w=M}, at the crossing, the eigenvectors $\ket{\psi_0}$ and $\ket{\psi_1}$ are each 50\% $\ket{s}$, 40\% $\ket{a}$, and 10\% $\ket{c}$. So, we expect that at the critical jumping rate, the system evolves from the initial state $\ket{s}$ to a final state that is 80\% $\ket{a}$ and 20\% $\ket{c}$.

Recall \cite{Wang2020} used degenerate perturbation theory to derive the jumping rate $\gamma_\text{\tiny WWW}$ \eqref{eq:gamma-WWW}, but it did not lead to a derivation of the evolution. Instead, it was numerically shown to yield a success probability of 0.36. This is because $\gamma_\text{\tiny WWW}$ is not precise enough. With $M = 10\,000$, it yields $\gamma M \approx 0.6667$, which is slightly too small compared to the actual 0.6668 in \fref{fig:overlaps-w=M}. Here, we will take into account which edges are most plentiful when we select the terms in the leading-order Hamiltonian, leading to a more precise critical jumping rate, and we will use a change of basis \cite{Wong5,Wong20,Wong29} to derive the evolution, showing that the success probability reaches 0.8. Beginning with the change of basis, instead of using the ``unprimed'' basis states $\{\ket{a},\ket{b},\ket{c},\ket{d},\ket{e},\ket{f},\ket{g}\}$, we use the following ``primed'' basis:
\begin{align*}
    & \ket{a'} = \ket{a}, \\
    & \ket{b'} = \frac{1}{\sqrt{M-1}} \left( -\sqrt{M-2}\ket{b} + \ket{g} \right), \\
    & \ket{c'} = \ket{c}, \\
    & \ket{d'} = \frac{1}{\sqrt{M(M-1)}} \left( \ket{b} - (M-1) \ket{d} + \sqrt{M-2} \ket{g} \right), \\
    & \ket{e'} = \ket{e}, \\
    & \ket{f'} = \frac{1}{\sqrt{M(M+1)}} \left(\ket{b} + \ket{d} - M\ket{f} + \sqrt{M-2}\ket{g}\right), \\
    & \ket{g'} = \frac{1}{\sqrt{M+1}} \left(\ket{b} + \ket{d} + \ket{f} + \sqrt{M-2}\ket{g}\right).
\end{align*}
Note that each primed basis state is asymptotically equal to its unprimed version, e.g., $\ket{b'} \approx \ket{b}$ up to a phase for large $M$. We came upon this basis by trial and error, but an intuition that guided us was that we needed $\ket{g'}$, which is approximately the starting state, to contain some $\ket{b}$ and $\ket{d}$. Since the $b$ vertices are adjacent to the $a$ vertex, and the $d$ vertices are adjacent to the $c$ vertex, this allows the math to capture the evolution from $\ket{g'}$ to a combination of $\ket{a}$ and $\ket{c}$.

To convert the Hamiltonian from the unprimed basis to the primed basis, we compute $P^{-1}HP$, where
\[ 
    P = \begin{pmatrix}
        \ket{a'} & \ket{b'} & \ket{c'} & \ket{d'} & \ket{e'} & \ket{f'} & \ket{g'}
    \end{pmatrix}.
\]
This produces $H$ in a new basis, which we call $H'$. To better gauge how the entries of the matrix scale relative to each other, we write $H'$ up to constant order and simplify terms by letting $w=M$:
\[ H' \approx -\gamma \begin{pmatrix}
    \frac{1}{\gamma} & -\sqrt{M} & M & 0 & 0 & 0 & 1 \\
    -\sqrt{M} & M-1 & 0 & 1 & \frac{3}{2} - M & 0 & \sqrt{M} \\
    M & 0 & 0 & -\sqrt{M} & 0 & 0 & 1 \\
    0 & 1 & -\sqrt{M} & M-1 & 2 & M-1 & 0 \\
    0 & \frac{3}{2} - M & 0 & 2 & 0 & 1 & 2 \sqrt{M} \\
    0 & 0 & 0 & M-1 & 1 & -2 & 0 \\
    1 & \sqrt{M} & 1 & 0 & 2 \sqrt{M} & 0 & 2M-3
\end{pmatrix}. \]
Now, we have edges connecting $\ket{a}$ to $\ket{g'}$ and $\ket{c}$ to $\ket{g'}$, so we can proceed with the perturbative calculation. $H'$ can be described as the sum of three matrices, $H'^{(0)},H'^{(1)},H'^{(2)}$, where apart from an overall factor of $-\gamma$, the terms of each matrix scale with $M, \sqrt{M}$, and constants, respectively:
\begin{gather*}
    H'^{(0)} = -\gamma \begin{pmatrix}
        \frac{1}{\gamma} & 0 & M & 0 & 0 & 0 & 0 \\
        0 &  M & 0 & 0 & - M & 0 & 0 \\
        M & 0 & 0 & 0 & 0 & 0 & 0 \\
        0 & 0 & 0 &  M & 0 & M & 0 \\
        0 & - M & 0 & 0 & 0 & 0 & 0 \\
        0 & 0 & 0 &  M & 0 & 0 & 0 \\
        0 & 0 & 0 & 0 & 0 & 0 & 2M-3
    \end{pmatrix}, \\
    H'^{(1)} = -\gamma \begin{pmatrix}
        0 & -\sqrt{M} & 0 & 0 & 0 & 0 & 0 \\
        -\sqrt{M} & 0 & 0 & 0 & 0 & 0 & \sqrt{M} \\
        0 & 0 & 0 & -\sqrt{M} & 0 & 0 & 0 \\
        0 & 0 & -\sqrt{M} & 0 & 0 & 0 & 0 \\
        0 & 0 & 0 & 0 & 0 & 0 & 2 \sqrt{M} \\
        0 & 0 & 0 & 0 & 0 & 0 & 0 \\
        0 & \sqrt{M} & 0 & 0 & 2 \sqrt{M} & 0 & 0
    \end{pmatrix}, \\
    H'^{(2)} = -\gamma \begin{pmatrix}
        0 & 0 & 0 & 0 & 0 & 0 & 1 \\
        0 & -1 & 0 & 1 & \frac{3}{2} & 0 & 0 \\
        0 & 0 & 0 & 0 & 0 & 0 & 1 \\
        0 & 1 & 0 & -1 & 2 & -1 & 0 \\
        0 & \frac{3}{2} & 0 & 2 & 0 & 1 & 0 \\
        0 & 0 & 0 & -1 & 1 & -2 & 0 \\
        1 & 0 & 1 & 0 & 0 & 0 & 0
    \end{pmatrix}.
\end{gather*}
Note that in $H'^{(0)}$, we kept a constant of $-3$ in the bottom-right corner because it comes from the edges connecting $g$ vertices to $g$ vertices, and from \tref{table:edges}, these are the most plentiful kinds of edges, and following \cite{Wong16}, this constant plays a greater role than other constants. Without this, the critical jumping rate we would derive would not be precise enough.

\begin{table}
    \caption{\label{table:edges}The number of edges of each type in the simplex of complete graphs, adapted from Tables~I and II from \cite{Wong16}.}
    \begin{center}
    \begin{tabular}{ccc}
        \toprule
        Connection & No.~of Weight 1 Edges & No.~of Weight $w$ Edges \\
        \midrule
        $a \sim b$ & $M-1$ & 0 \\
        $a \sim c$ & 0 & $1$ \\
        $b \sim b$ & $\frac{1}{2}(M-1)(M-2)$ & 0 \\
        $b \sim e$ & 0 & $M - 1$ \\
        $c \sim d$ & $M-1$ & 0 \\
        $d \sim d$ & $\frac{1}{2}(M-1)(M-2)$ & 0 \\
        $d \sim f$ & 0 & $M - 1$ \\
        $e \sim f$ & $M - 1$ & 0 \\
        $e \sim g$ & $(M-1)(M-2)$ & 0 \\
        $f \sim g$ & $(M-1)(M-2)$ & 0 \\
        $g \sim g$ & $\frac{1}{2}(M-1)(M-2)(M-3)$ & $\frac{1}{2}(M-1)(M-2)$ \\
        \midrule
        Total & $\frac{1}{2}M(M^2-1)$ & $\frac{1}{2}M(M+1)$ \\
        \bottomrule
    \end{tabular}        
    \end{center}
\end{table}

\begin{table}
\begin{center}
\caption{\label{table:gammas}Different weights $w$ and their corresponding critical jumping rates.}
\begin{tabular}{ccc}
    \toprule
    Weight & Critical Jumping Rate & Eq. \\
    \midrule
    $w = M^{1/4}$ & $\displaystyle \frac{1}{M} + \frac{1}{M^{5/4}} - \frac{1}{M^{7/4}} - \frac{3}{M^{9/4}} + \frac{2}{M^{10/4}}$ & \eqref{eq:gamma-w=M0.25-W} \\
    \midrule
    $w=\sqrt{M}$ & $\displaystyle \frac{1}{M-2}$ & \eqref{eq:gamma-w=M0.5} \\
    \midrule
    $w = M^{5/8}$ & $\displaystyle \frac{1}{M} - \frac{1}{M^{11/8}} + \frac{1}{M^{13/8}} + \frac{2}{M^{7/4}} - \frac{4}{M^{17/8}}$ & \eqref{eq:gamma-w=M0.625} \\
    \midrule
    $w = M^{3/4}$ & $\displaystyle \frac{1}{M}-\frac{1}{M^{5/4}}+\frac{2}{M^{3/2}}-\frac{3}{M^{7/4}}+\frac{8}{M^{2}}-\frac{14}{M^{9/4}}+\frac{28}{M^{5/2}}-\frac{58}{M^{11/4}}+\frac{103}{M^{3}}$ & \eqref{eq:gamma-a-w=M0.75} \\
    \midrule
    $w = M^{7/8}$ & $\begin{aligned}
        &\frac{1}{M} - \frac{1}{M^{9/8}} + \frac{2}{M^{5/4}} - \frac{4}{M^{11/8}} + \frac{8}{M^{3/2}} - \frac{16}{M^{13/8}} \\
        &\quad+ \frac{32}{M^{7/4}} - \frac{63}{M^{15/8}} + \frac{128}{M^2} - \frac{254}{M^{17/8}} + \frac{512}{M^{9/4}} - \frac{1079}{M^{19/8}} \\
        &\quad+ \frac{2864}{M^{5/2}} - \frac{12075}{M^{21/8}} + \frac{60988}{M^{11/4}} - \frac{229326}{M^{23/8}} + \frac{420554}{M^3} \\
    \end{aligned}$ & \eqref{eq:gamma-w=M0.875} \\
    \midrule
    $w=M$ & $\displaystyle \frac{2M-3}{3(M-1)(M-3)}$ & \eqref{eq:gamma-w=M} \\
    \midrule
    $w \gg M$ & $\displaystyle \frac{M+w-2}{(M-2)(M+2w-2)}$ & \eqref{eq:gamma-w>>M} \\
    \bottomrule
\end{tabular}
\end{center}
\end{table}

Now, we solve for the eigenvectors and eigenvalues of $H'^{(0)}$. The two relevant ones are
\begin{align*}
    &\ket{g'}, &&E_g=-\gamma(2M-3), \\
    &\frac{1+\sqrt{1+4\gamma^2M^2}}{2\gamma M}\ket{a}+\ket{c}, &&E_{ac}=-\frac{1+\sqrt{1+4\gamma^2M^2}}{2}.
\end{align*}
These eigenvectors are degenerate when $\gamma$ assumes the critical value of
\begin{equation}
    \label{eq:gamma-w=M}
    \gamma_{w=M} = \frac{2M-3}{3(M-1)(M-3)}.
\end{equation}
For example, when $M = 10\,000$, this yields $\gamma M \approx 0.6668$, in agreement with \fref{fig:overlaps-w=M}. This critical jumping rate is listed in \tref{table:gammas}, along with others from the rest of the paper. At this critical jumping rate, the above two eigenvectors and eigenvalues of $H'^{(0)}$ become
\begin{align*}
    &\ket{g'}, &&E_g = -\frac{(2M-3)^2}{3(M-1)(M-3)}, \\
    &\frac{2M-3}{M}\ket{a}+\ket{c} \rightarrow \ket{ac} = \frac{2}{\sqrt{5}} \ket{a} + \frac{1}{\sqrt{5}} \ket{c}, &&E_{ac} = -\frac{(2M-3)^2}{3(M-1)(M-3)},
\end{align*}
where we have normalized the second eigenvector for large $M$ and called it $\ket{ac}$.

At the critical jumping rate, since $\ket{g'}$ and $\ket{ac}$ have the same eigenvalue, any linear combination of them,
\begin{equation}
    \label{eq:linear-comb}
    \alpha_{ac} \ket{ac} + \alpha_g \ket{g},
\end{equation}
is also an eigenvector with the same eigenvalue. In degenerate perturbation theory \cite{Griffiths2005}, we add a perturbation $H^{(1)}$ and look for eigenvectors of $H^{(0)} + H^{(1)}$ that take the form \eqref{eq:linear-comb}. That is, we look for solutions to the eigenvalue relation
\[ \left( H'^{(0)}+H'^{(1)} \right) \left( \alpha_{ac} \ket{ac} + \alpha_g \ket{g} \right) = E \left( \alpha_{ac} \ket{ac} + \alpha_g \ket{g} \right), \]
where $E$ is the eigenvalue. This does not perturb the two degenerate eigenvectors, however. That is, $\ket{g'}$ and $\ket{ac}$ remain degenerate eigenvectors of $H^{(0)} + H^{(1)}$. Rather, $H'^{(2)}$ behaves as the first-order correction to the Hamiltonian. So, we look for eigenvectors of $H'^{(0)}+H'^{(1)}+H'^{(2)}$ of the form \eqref{eq:linear-comb}, whose coefficients and eigenvalues can be found by solving the eigenvalue relation
\[ \left( H'^{(0)}+H'^{(1)}+H'^{(2)} \right) \left( \alpha_{ac} \ket{ac} + \alpha_g \ket{g} \right) = E \left( \alpha_{ac} \ket{ac} + \alpha_g \ket{g} \right). \]
In matrix-vector form, this eigenvalue relation is
\[ \begin{pmatrix}
    \frac{(2M-3)^2}{3(M-1)(M-3)} & \frac{2M-3}{(M-3) \sqrt{5M^2 - 12M + 9}} \\
    \frac{2M-3}{(M-3) \sqrt{5M^2 - 12M + 9}} & \frac{(2M-3)^2}{3(M-1)(M-3)}
\end{pmatrix}
\begin{pmatrix}
    \alpha_{ac}\\ \alpha_{g}
\end{pmatrix}=E\begin{pmatrix}
    \alpha_{ac}\\ \alpha_{g}
\end{pmatrix}. \]
Solving this, we get the following two eigenvectors and eigenvalues of $H'^{(0)}+H'^{(1)}+H'^{(2)}$:
\begin{align*}
    &\ket{\psi_0}= \frac{1}{\sqrt{2}}\left(\ket{ac}+\ket{g}\right), && E_0 = -\frac{(2M-3) \left[ (2M-3)R + 3(M-1) \right]}{3(M-1)(M-3)R}, \\
    &\ket{\psi_1}= \frac{1}{\sqrt{2}}\left(\ket{ac}-\ket{g}\right), && E_1 = -\frac{(2M-3) \left[ (2M-3)R - 3(M-1) \right]}{3(M-1)(M-3)R},
\end{align*}
where $R=\sqrt{5M^2 - 12M + 9}$. For large $M$, $R \approx \sqrt{5}M$ and $\Delta E \approx 4/R = 4/(\sqrt{5} M)$, and so the system evolves from $\ket{s} \approx \ket{g'}$ to $\ket{ac}$ in time
\begin{equation}
    \label{eq:t-w=M}
    t_{w=M} = \frac{\pi}{\Delta E} = \frac{\pi\sqrt{5}}{4}M = \frac{\pi\sqrt{5}}{4}\sqrt{N},
\end{equation}
for large $M$ or $N$. Since $\ket{ac} = (2\ket{a} + \ket{c})/\sqrt{5}$, measuring the position of the walker at this time results in $\ket{a}$ with probability 0.8, or $\ket{c}$ with probability 0.2, completing the proof of the $w = M$ result in \fref{fig:summary}.

\begin{SCfigure}
    \caption{\label{fig:prob-time-w=M}For search on the simplex of complete graphs with $w = M = 10\,000$ and $\gamma = \gamma_{w=M}$ \eqref{eq:gamma-w=M}, the probability at the marked vertex $\ket{a}$ (solid black) and at vertex $\ket{c}$ (dashed red).}
    \includegraphics[width=2.25in]{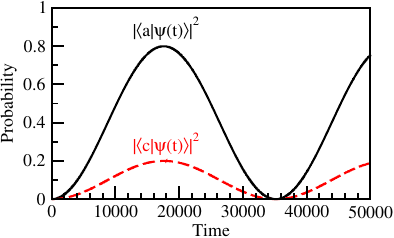}
\end{SCfigure}

\begin{table}
    \caption{\label{table:peak-success-runtime-w=M}For search on the simplex of complete graphs with bridge weights $w = M$ and jumping rate $\gamma = \gamma_{w = M}$ \eqref{eq:gamma-w=M}, numerically determined probabilities in $\ket{a}$ and $\ket{c}$ at time $t_{w = M}$ \eqref{eq:t-w=M}.}
    \begin{center}
    \begin{tabular}{ccc}
        \toprule
        $M$ & Prob. in $\ket{a}$ at $t_{w = M}$ & Prob. in $\ket{c}$ at $t_{w = M}$ \\
        \midrule
        $10^5$    & 0.7999530790619019493 & 0.19999692606744294660 \\
        $10^6$    & 0.7999955554622798069 & 0.19999944456820016430 \\
        $10^7$    & 0.7999995430322726822 & 0.19999995696800208596 \\
        $10^8$    & 0.7999999550716057265 & 0.19999999492840037359 \\
        $10^9$    & 0.7999999957276694267 & 0.19999999927233056581 \\
        $10^{10}$ & 0.7999999995649255883 & 0.19999999993507441213 \\
        \bottomrule
    \end{tabular}
    \end{center}
\end{table}

As a check, in \fref{fig:prob-time-w=M}, we plot the probability in $\ket{a}$ and $\ket{c}$ as the system evolves in time with $M = 10\,000$. At $t_{w=M} = 17\,562.9$, the probability at vertex $\ket{a}$ peaks around 80\%, and the probability at vertex $\ket{c}$ peaks around 20\%. As another check, we can similarly find the probability at $\ket{a}$ and $\ket{c}$ as $M$ increases. This is shown in \tref{table:peak-success-runtime-w=M}, and as expected, the probability at $\ket{a}$ converges to 80\%, and the probability at $\ket{c}$ converges to 20\%.

As discussed above, picking the right basis for degenerate perturbation theory to work was a matter of trial and error. As such, we were unsuccessful in making the approach work for the remaining weights, so the sections below are numerical rather than analytical.


\section{\label{sec:w=M0.5}Search when \texorpdfstring{$w = \sqrt{M}$}{w = sqrt(M)}}

\begin{figure}
\begin{center}
    \subfloat[] {
        \includegraphics[width=2.25in]{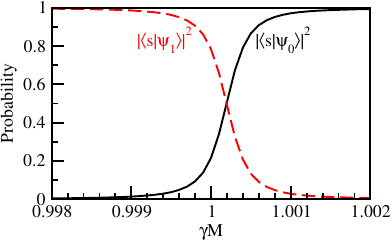}
    }
    \subfloat[] {
        \includegraphics[width=2.25in]{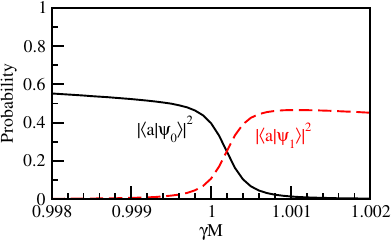}
    }

    \subfloat[] {
        \includegraphics[width=2.25in]{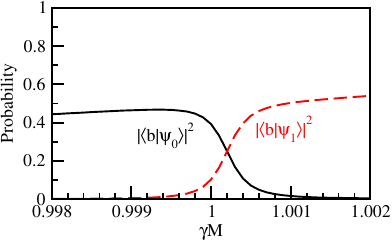}
    }
    \caption{\label{fig:overlaps-w=M0.5}For the simplex of complete graphs with $M = 10\,000$ and $w = \sqrt{M}$, probability overlaps of (a) the initial state $\ket{s}$, (b) the marked vertex $\ket{a}$, and (c) $\ket{b}$ with two eigenvectors $\ket{\psi_0}$ (solid black) and $\ket{\psi_1}$ (dashed red) of the search Hamiltonian for various values of $\gamma M$.}
\end{center}
\end{figure}

We next consider when $w = \sqrt{M}$ exactly. As before, we begin by calculating the probability overlaps of the eigenvectors of the search Hamiltonian \eqref{eq:H-7D} with the initial state $\ket{s}$, the marked vertex $\ket{a}$, and $\ket{b}$, now shown in \fref{fig:overlaps-w=M0.5}. Now, the critical jumping rate occurs around $\gamma M = 1.0002$, and the eigenvectors $\ket{\psi_0}$ and $\ket{\psi_1}$ are each half $\ket{s}$, a quarter $\ket{a}$, and a quarter $\ket{b}$, and so we expect that for this jumping rate, the system evolves from the initial state $\ket{s}$ to equal parts $\ket{a}$ and $\ket{b}$.

\begin{SCfigure}
    \caption{\label{fig:prob-time-w=M0.5}For search on the simplex of complete graphs with $M = 10\,000$, $w = \sqrt{M}$, and $\gamma = \gamma_{w = \sqrt{M}}$ \eqref{eq:gamma-w=M0.5}, the probability at the marked vertex $\ket{a}$ (solid black) and at vertex $\ket{b}$ (dashed red).}
    \includegraphics[width=2.25in]{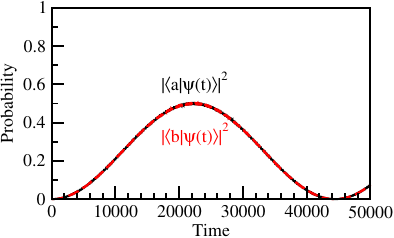}
\end{SCfigure}

We observe from \fref{fig:overlaps-w=M0.5} that the critical jumping rate occurs just above $\gamma M = 1$, or $\gamma = 1/M$. We guessed the following critical jumping rate,
\begin{equation}
    \label{eq:gamma-w=M0.5}
    \gamma_{w = \sqrt{M}} = \frac{1}{M-2},
\end{equation}
which is slightly larger than $1/M$, and numerically confirmed that it works for various large $M$. For example, when $M = 10\,000$, this yields $\gamma \approx 1.0002$, in agreement with \fref{fig:overlaps-w=M0.5}. As a check, using these parameters, we plot in \fref{fig:prob-time-w=M0.5} the probability in $\ket{a}$ and $\ket{b}$ as the stystem evolves with time, and they each reach roughly 50\%, as expected from \fref{fig:overlaps-w=M0.5}.

\begin{table}
    \caption{\label{table:peak-success-runtime-w=M0.5}For search on the simplex of complete graphs with bridge weights $w = \sqrt{M}$ and jumping rate $\gamma = \gamma_{w=\sqrt{M}}$ \eqref{eq:gamma-w=M0.5}, numerically determined probabilities in $\ket{a}$ and $\ket{b}$ at time $t_{w = \sqrt{N}}$ \eqref{eq:t-w=M0.5}.}
    \begin{center}
    \begin{tabular}{ccc}
        \toprule
        $M$ & Prob. in $\ket{a}$ at $t_{w=\sqrt{M}}$ & Prob. in $\ket{b}$ at $t_{w=\sqrt{M}}$ \\
        \midrule
        $10^5$ & 0.49927352714560 & 0.50070366154557 \\
        $10^6$ & 0.50022079888708 & 0.49977692006982 \\
        $10^7$ & 0.50004644750028 & 0.49995332438040 \\
        $10^8$ & 0.50001462569395 & 0.49998535149373 \\
        $10^9$ & 0.49998988825619 & 0.50001010946256 \\
        $10^{10}$ & 0.49999681874746 & 0.50000318102441 \\
        \bottomrule
    \end{tabular}
    \end{center}
\end{table}

The time at which the probabilities peak in \fref{fig:prob-time-w=M0.5} is roughly
\begin{equation}
    \label{eq:t-w=M0.5}
    t_{w = \sqrt{M}} = \frac{\pi}{\sqrt{2}} \sqrt{N},
\end{equation}
which with $M = 10\,000$ is $2.22 \times 10^4$. In \tref{table:peak-success-runtime-w=M0.5}, we similarly find the probability in $\ket{a}$ and $\ket{b}$ at this runtime for various sizes of graphs. We see that the probability in each converges to 50\% for large $N$. Thus, we numerically conclude that when $w = \sqrt{M}$ and at the critical jumping rate \eqref{eq:gamma-w=M0.5}, the walker evolves from $\ket{s}$ to half $\ket{a}$ and half $\ket{b}$ in time \eqref{eq:t-w=M0.5}, as mentioned in the introduction and noted in \fref{fig:summary}.


\section{\label{sec:w<<M^0.5}Search when \texorpdfstring{$w \ll \sqrt{M}$}{w << sqrt(M)}}

In this section, we consider when $w$ scales less than $\sqrt{M}$. From \cite{Wong16}, we expect that the system evolves to $\ket{b}$, but here we will see that a more precise jumping rate may be needed. To illustrate this, we give an example below with $w = M^{1/4}$.


\subsection{Example: $w = M^{1/4}$}

\begin{figure}
\begin{center}
    \subfloat[] {
        \includegraphics[width=2.25in]{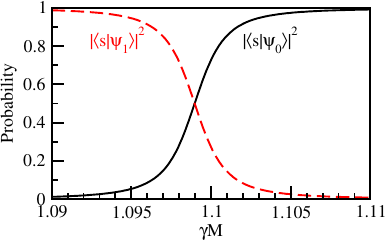}
        \label{fig:overlaps-w=M0.25-s}
    }
    \subfloat[] {
        \includegraphics[width=2.25in]{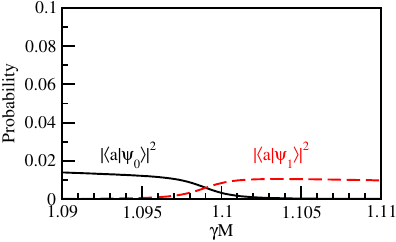}
        \label{fig:overlaps-w=M0.25-a}
    }
    
    \subfloat[] {
        \includegraphics[width=2.25in]{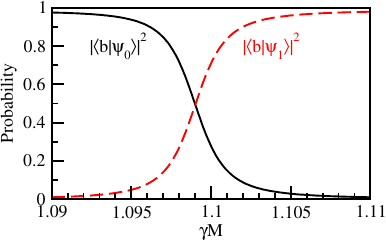}
        \label{fig:overlaps-w=M0.25-b}
    }
    \caption{\label{fig:overlaps-w=M0.25}For the simplex of complete graphs with $M = 10\,000$ and $w = M^{1/4}$, probability overlaps of (a) the initial state $\ket{s}$, (b) the marked vertex $\ket{a}$, and (c) $\ket{b}$ with two eigenvectors $\ket{\psi_0}$ (solid black) and $\ket{\psi_1}$ (dashed red) of the search Hamiltonian for various values of $\gamma M$.}
\end{center}
\end{figure}

For our example, we take $w = M^{1/4}$, which falls within the regime analyzed by \cite{Wong16}. As before, in \fref{fig:overlaps-w=M0.25}, we plot the probability overlaps of two eigenvectors of the search Hamiltonian \eqref{eq:H-7D} with $\ket{s}$, $\ket{a}$, and $\ket{b}$. We observe that the critical jumping rate occurs around $\gamma M = 1.099$. At the crossing, the eigenvectors $\ket{\psi_0}$ and $\ket{\psi_1}$ are each half $\ket{s}$, while the remaining half is mostly $\ket{b}$ with a tiny bit of $\ket{a}$. Then, at the critical jumping rate, the system evolves from the initial state $\ket{s}$ to mostly $\ket{b}$ and a tiny amount $\ket{a}$. As $M$ increases, the contribution in $\ket{b}$ goes to 1, and the contribution in $\ket{a}$ goes to 0, which is the asymptotic result from \cite{Wong16}.

As discussed in the introduction, \cite{Wong16} showed that the time it takes for the system to evolve from $\ket{s}$ to asymptotically $\ket{b}$ is
\begin{equation}
    \label{eq:t-w<M^0.5}
    t_{w \ll \sqrt{N}} = \frac{\pi}{2(1+w)} N^{3/4}.
\end{equation}
Using degenerate perturbation theory, \cite{Wong16} proposed that the critical jumping rate that accomplishes this is $(1 + 1/w)/M$, which for $w = M^{1/4}$ yields
\begin{equation}
    \label{eq:gamma-w=M0.25-W}
    \gamma_\text{\tiny W} = \frac{1}{M} + \frac{1}{M^{5/4}}.
\end{equation}
When $w = M^{1/4}$ and $M = 10\,000$, this yields $\gamma_\text{\tiny W} M = 1.1$, which from \fref{fig:overlaps-w=M0.25} is slightly too big, as the actual value should be around 1.099.

\begin{table}
    \caption{\label{table:gamma-data-w=M0.25}For search on the simplex of complete graphs with bridge weights $w = M^{1/4}$, numerically determined values of the critical jumping rate $\gamma$ as a function of $M$.}
    \begin{center}
    \begin{tabular}{cc}
        \toprule
        $M$ & Critical $\gamma$ \\
        \midrule
        $10^{8}$  & $1.009998999702029599920584165918507981009 \times 10^{-8\phantom{0}}$ \\
        $10^{9}$  & $1.005623235407156022192729825738133991606 \times 10^{-9\phantom{0}}$ \\
        $10^{10}$ & $1.003162246036445103796987485882667117496 \times 10^{-10}$ \\
        $10^{11}$ & $1.001778273786572385929288178352338270051 \times 10^{-11}$ \\
        $10^{12}$ & $1.000999998999997002002996136616411274561 \times 10^{-12}$ \\
        $10^{13}$ & $1.000562341147362239437404013730336826455 \times 10^{-13}$ \\
        $10^{14}$ & $1.000316227734394051846684065422712770187 \times 10^{-14}$ \\
        $10^{15}$ & $1.000177827935380478494798490877128515400 \times 10^{-15}$ \\
        $10^{16}$ & $1.000099999998999999970002000418106635613 \times 10^{-16}$ \\
        $10^{17}$ & $1.000056234132341206965348642160966473080 \times 10^{-17}$ \\
        $10^{18}$ & $1.000031622776570061016623438805181089538 \times 10^{-18}$ \\
        $10^{19}$ & $1.000017782794094765814755015947599187300 \times 10^{-19}$ \\
        $10^{20}$ & $1.000009999999998999999999700002087644994 \times 10^{-20}$ \\
        \bottomrule
    \end{tabular}
    \end{center}
\end{table}

\begin{figure}
\begin{center}
    \subfloat[$\gamma = \gamma_{w = M^{1/4}}$]{
        \includegraphics[width=2.25in]{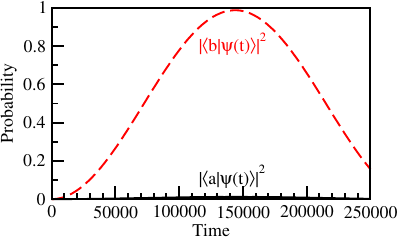}
        \label{fig:prob-time-w=M0.25-num}
        
    }
    \subfloat[$\gamma = \gamma_\text{\tiny W}$]{
        \includegraphics[width=2.25in]{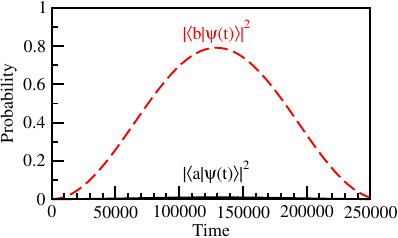}
        \label{fig:prob-time-w=M0.25-gammaW}
    }
    
    \caption{\label{fig:prob-time-w=M0.25}For search on the simplex of complete graphs with $M = 10\,000$ and  $w = M^{1/4}$, the probability at the marked vertex $\ket{a}$ (solid black) and at vertex $\ket{b}$ (dashed red) with (a) $\gamma = \gamma_{w = M{1/4}}$ \eqref{eq:gamma-w=M0.25} and (b) $\gamma = \gamma_\text{\tiny W}$ \eqref{eq:gamma-w=M0.25-W}.}
\end{center}
\end{figure}

Thus, in this paper, we observe that a more precise critical jumping rate is needed. Numerically, we can determine it by finding where $|\braket{s}{\psi_0}|^2 = |\braket{s}{\psi_1}|^2$, i.e., where the curves intersect in \fref{fig:overlaps-w=M0.25-s}. Doing this for various values of $M$, we get the values shown in Table \ref{table:gamma-data-w=M0.25}. Then, we fit the following function of $M$ to it:
\[ \frac{a_1}{M} + \frac{a_2}{M^{5/4}} + \frac{a_3}{M^{7/4}} + \frac{a_4}{M^{9/4}} + \frac{a_5}{M^{10/4}}, \]
where the $a_i$'s are constants. Doing the fit, we obtain the following critical jumping rate when $w = M^{1/4}$:
\begin{equation}
    \label{eq:gamma-w=M0.25}
    \gamma_{w = M^{1/4}} = \frac{1}{M} + \frac{1}{M^{5/4}} - \frac{1}{M^{7/4}} - \frac{3}{M^{9/4}} + \frac{2}{M^{10/4}}.
\end{equation}
For example, when $M = 10\,000$, $\gamma_{w = M^{1/4}} M \approx 1.099$, as expected from \fref{fig:overlaps-w=M0.25}. In \fref{fig:prob-time-w=M0.25}, the probability of measuring the walker at $\ket{a}$ or $\ket{b}$ as a function of time is plotted. In \fref{fig:prob-time-w=M0.25-num}, $\gamma_{w = M^{1/4}}$ is used, and almost all of the probability accumulates in $\ket{b}$. In contrast, in \fref{fig:prob-time-w=M0.25-gammaW}, $\gamma_\text{\tiny W}$ is used, and it is not accurate enough, so the probability is not entirely contained in $\ket{a}$ and $\ket{b}$.

\begin{table}
    \caption{\label{table:peak-success-runtime-w=M0.25}For search on the simplex of complete graphs with bridge weights $w = M^{1/4}$ and jumping rate $\gamma = \gamma_{w=M{1/4}}$ \eqref{eq:gamma-w=M0.25}, numerically determined probabilities in $\ket{a}$ and $\ket{b}$ at time $t_{w \ll \sqrt{N}}$ \eqref{eq:t-w<M^0.5}.}
    \begin{center}
    \begin{tabular}{ccc}
        \toprule
        $M$ & Prob. in $\ket{a}$ at $t_{w \ll \sqrt{M}}$ & Prob. in $\ket{b}$ at $t_{w \ll \sqrt{M}}$ \\
        \midrule
        $10^5$    & 0.003494532554548757056 & 0.9964864783640590811 \\
        $10^6$    & 0.001060998423027477617 & 0.9989372262792127317 \\
        $10^7$    & 0.000327352449728619326 & 0.9996724769964047541 \\
        $10^8$    & 0.000102007050023686877 & 0.9998979762847201910 \\
        $10^9$    & 0.000031978232269486727 & 0.9999680201230205402 \\
        $10^{10}$ & 0.000010063438803720463 & 0.9999899363979471722 \\
        \bottomrule
    \end{tabular}
    \end{center}
\end{table}

While the critical jumping rate derived in \cite{Wong16} was not precise enough, the runtime \eqref{eq:t-w<M^0.5} is. Evaluating this runtime with $w = M^{1/4}$ and $M = 10\,000$ yields $142\,810$, which agrees with \fref{fig:overlaps-w=M0.25}. As further confirmation of the runtime's correctness, in \tref{table:peak-success-runtime-w=M0.25}, we numerically find the probability in $\ket{a}$ and $\ket{b}$ at the runtime for various values of $M$, and we see that the probability in $\ket{a}$ goes to 0 while the probability in $\ket{b}$ goes to 1, in agreement with \cite{Wong16}. This result is reflected in the bottom-left corner of \fref{fig:summary}.


\section{\label{sec:M^0.5<<w<<M}Search when \texorpdfstring{$\sqrt{M} \ll w \ll M$}{sqrt(M) << w << M}}

In this section, we consider when $w$ scales between $\sqrt{M}$ and $M$. As discussed in the introduction and summarized in \fref{fig:summary}, for large $M$, the system should evolve to just $\ket{a}$ in time $\pi\sqrt{N}/2$.

In the following three subsections, we will give three examples of $w$ in this region, which will be when $w = M^{5/8}$, $M^{3/4}$, and $M^{7/8}$.


\subsection{Example: $w = M^{5/8}$}

For our first example where $\sqrt{M} \ll w \ll M$, we consider $w = M^{5/8}$.

\begin{figure}
\begin{center}
    \subfloat[] {
        \includegraphics[width=2.25in]{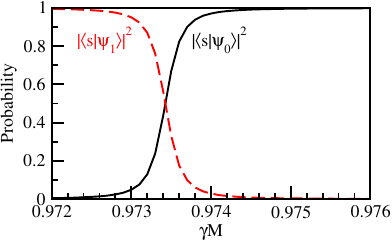}
        \label{fig:overlaps-w=M0.625-s}
    }
    \subfloat[] {
        \includegraphics[width=2.25in]{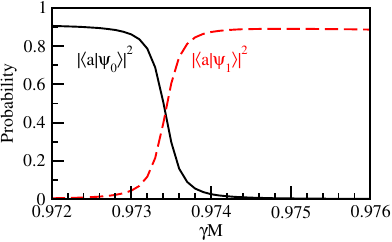}
    }

    \subfloat[] {
        \includegraphics[width=2.25in]{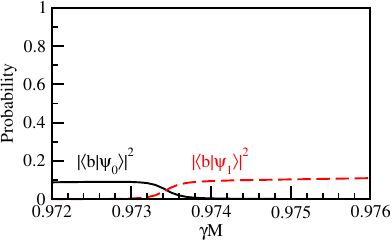}
    }
    \caption{\label{fig:overlaps-w=M0.625}For the simplex of complete graphs with $M = 10\,000$ and $w = M^{5/8}$, probability overlaps of (a) the initial state $\ket{s}$, (b) the marked vertex $\ket{a}$, and (c) $\ket{b}$ with two eigenvectors $\ket{\psi_0}$ (solid black) and $\ket{\psi_1}$ (dashed red) of the search Hamiltonian for various values of $\gamma M$.}
\end{center}
\end{figure}

In \fref{fig:overlaps-w=M0.625}, the probability overlaps of two eigenvectors of the search Hamiltonian with $\ket{s}$, $\ket{a}$, and $\ket{b}$ are shown when $M = 10\,000$. At the critical jumping rate, $\gamma M \approx 0.9734$, the eigenvectors $\ket{\psi_0}$ and $\ket{\psi_1}$ are both half $\ket{s}$, while the remaining half is mostly $\ket{a}$ with a little $\ket{b}$. So, the system evolves from $\ket{s}$ to mostly $\ket{a}$ and a little $\ket{b}$. As $M$ increases, more and more of the final state is in $\ket{a}$, and less and less of it is in $\ket{b}$.

\begin{table}
    \caption{\label{table:gamma-data-w=M0.625}For search on the simplex of complete graphs with bridge weights $w = M^{5/8}$, numerically determined values of the critical jumping rate $\gamma$ as a function of $M$.}
    \begin{center}
    \begin{tabular}{cc}
        \toprule
        $M$ & Critical $\gamma$ \\
        \midrule
        $10^{8}$  & $9.990119960277427176525652142450311049102 \times 10^{-9\phantom{0}}$ \\
        $10^{9}$  & $9.995810302272920947512263911510945423909 \times 10^{-10}$ \\
        $10^{10}$ & $9.998227976234242858426224435148856840808 \times 10^{-11}$ \\
        $10^{11}$ & $9.999251551763514383809212739023757474552 \times 10^{-12}$ \\
        $10^{12}$ & $9.999684108460484979618329714136056942791 \times 10^{-13}$ \\
        $10^{13}$ & $9.999866726402668595157103783311145534303 \times 10^{-14}$ \\
        $10^{14}$ & $9.999943784282723485601537613419935853690 \times 10^{-15}$ \\
        $10^{15}$ & $9.999976290592376149412651543499477390874 \times 10^{-16}$ \\
        $10^{16}$ & $9.999990001019999960002079795829499340621 \times 10^{-17}$ \\
        $10^{17}$ & $9.999995783275659640564267708610874382670 \times 10^{-18}$ \\
        $10^{18}$ & $9.999998221777456548903334464853618317227 \times 10^{-19}$ \\
        $10^{19}$ & $9.999999250119238350113976300257393592636 \times 10^{-20}$ \\
        $10^{20}$ & $9.999999683775416260820970274783068768213 \times 10^{-21}$ \\
        \bottomrule
    \end{tabular}
    \end{center}
\end{table}

\begin{SCfigure}
    \caption{\label{fig:prob-time-w=M0.625}For search on the simplex of complete graphs with $M = 10\,000$, $w = M^{5/8}$, and $\gamma = \gamma_{w = M^{5/8}}$ \eqref{eq:gamma-w=M0.625}, the probability at the marked vertex $\ket{a}$ (solid black) and at vertex $\ket{b}$ (dashed red).}
    \includegraphics[width=2.25in]{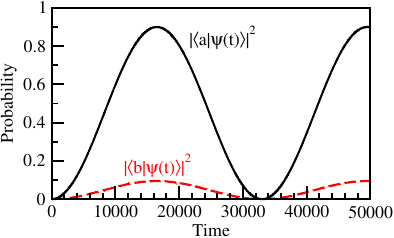}
\end{SCfigure}

For various values of $M$, we numerically find the critical jumping rate by finding where $|\braket{s}{\psi_0}|^2 = |\braket{s}{\psi_1}|^2$, i.e., where the curves intersect in \fref{fig:overlaps-w=M0.625-s}. These values are shown in Table \ref{table:gamma-data-w=M0.625}. Then, we fit the following function of $M$ to it:
\[ \frac{a_1}{M} + \frac{a_2}{M^{11/8}} + \frac{a_3}{M^{13/8}} + \frac{a_4}{M^{7/4}} + \frac{a_5}{M^{17/8}}, \]
where the $a_i$'s are constants. Doing the fit, we obtain the following critical jumping rate when $w = M^{5/8}$:
\begin{equation}
    \label{eq:gamma-w=M0.625}
    \gamma_{w=M{5/8}} = \frac{1}{M} - \frac{1}{M^{11/8}} + \frac{1}{M^{13/8}} + \frac{2}{M^{7/4}} - \frac{4}{M^{17/8}}.
\end{equation}
For example, when $M = 10\,000$, this yields $\gamma_{w = M^{5/8}} M = 0.9734$, as expected from \fref{fig:overlaps-w=M0.625}.

\begin{table}
    \caption{\label{table:peak-success-runtime-w=M0.625}For search on the simplex of complete graphs with bridge weights $w = M^{5/8}$ and jumping rate $\gamma = \gamma_{w=M{5/8}}$ \eqref{eq:gamma-w=M0.625}, numerically determined probabilities in $\ket{a}$ and $\ket{b}$ at time $t_{w = M^{5/8}}$ \eqref{eq:t-w=M0.625}.}
    \begin{center}
    \begin{tabular}{ccc}
        \toprule
        $M$ & Prob. in $\ket{a}$ at $t_{w=M{5/8}}$ & Prob. in $\ket{b}$ at $t_{w=M{5/8}}$ \\
        \midrule
        $10^5$    & 0.9429742694468459890 & 0.05436890092652864312 \\
        $10^6$    & 0.9683359583752317547 & 0.03094409522402743560 \\
        $10^7$    & 0.9822428879575553474 & 0.017546933646849227841 \\
        $10^8$    & 0.9900162222848821904 & 0.009919835735005671273 \\
        $10^9$    & 0.9943836132608895290 & 0.005596535025685250228 \\
        $10^{10}$ & 0.9968403765367597338 & 0.003153400844150417692 \\
        \bottomrule
    \end{tabular}
    \end{center}
\end{table}

Using our critical jumping rate \eqref{eq:gamma-w=M0.625} with $M = 10\,000$, we plot in \fref{fig:prob-time-w=M0.625} the probability in $\ket{a}$ and $\ket{b}$ as the system evolves with time, and we see that they peak around $t = 1.57 \times 10^4$, with the walker mostly in $\ket{a}$ and a little in $\ket{b}$. Numerically, we see that the runtime is
\begin{equation}
    \label{eq:t-w=M0.625}
    t_{w = M^{5/8}} = \frac{\pi}{2} \sqrt{N}.
\end{equation}
Numerically, we can find the probability in $\ket{a}$ and $\ket{b}$ at this runtime \eqref{eq:t-w=M0.625} for various values of $M$. this is shown in \tref{table:peak-success-runtime-w=M0.625}, and we see that as $M$ increases, the system asymptotically evolves entirely to $\ket{a}$.


\subsection{Example: $w = M^{3/4}$}

For our second example where $\sqrt{M} \ll w \ll M$, we consider $w = M^{3/4}$.

\begin{figure}
\begin{center}
    \subfloat[] {
        \includegraphics[width=2.25in]{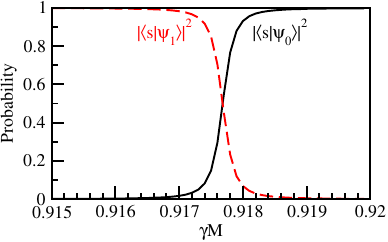}
        \label{fig:overlaps-w=M0.75-s}
    }
    \subfloat[] {
        \includegraphics[width=2.25in]{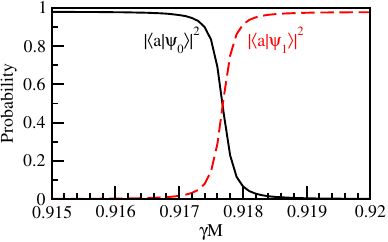}
    }
    \caption{\label{fig:overlaps-w=M0.75}For the simplex of complete graphs with $M = 10\,000$ and $w = M^{3/4}$, probability overlaps of (a) the initial state $\ket{s}$ and (b) the marked vertex $\ket{a}$ with two eigenvectors $\ket{\psi_0}$ (solid black) and $\ket{\psi_1}$ (dashed red) of the search Hamiltonian for various values of $\gamma M$.}
\end{center}
\end{figure}

As before, we begin by plotting the probability overlaps of some eigenvectors of the search Hamiltonian \eqref{eq:H-7D} with the starting state $\ket{s}$ and marked vertex $\ket{a}$, as shown in \fref{fig:overlaps-w=M0.75}. For the system to evolve apart from a global phase, the jumping rate must take a critical value where the curves cross, around $\gamma M \approx 0.9177$. At this crossing, the eigenvectors $\ket{\psi_0}$ and $\ket{\psi_1}$ are each half $\ket{s}$ and half $\ket{a}$, and so we expect that for this jumping rate, the system evolves from the initial state $\ket{s}$ to the marked vertex $\ket{a}$.

\begin{table}
    \caption{\label{table:gamma-data-w=M0.75}For search on the simplex of complete graphs with bridge weights $w = M^{3/4}$, numerically determined values of the critical jumping rate $\gamma$ as a function of $M$.}
    \begin{center}
    \begin{tabular}{cc}
        \toprule
        $M$ & Critical $\gamma$ \\
        \midrule
        $10^{8}$  & $9.901970786274305783520609904422965886876 \times 10^{-9\phantom{0}}$ \\
        $10^{9}$  & $9.944393067396243100464617755354827194871 \times 10^{-10}$ \\
        $10^{10}$ & $9.968576282671024445531785182894655646926 \times 10^{-11}$ \\
        $10^{11}$ & $9.982280283547935813164858263596864537843 \times 10^{-12}$ \\
        $10^{12}$ & $9.990019970079860279421077740693721454082 \times 10^{-13}$ \\
        $10^{13}$ & $9.994382905976570751903467634312257297501 \times 10^{-14}$ \\
        $10^{14}$ & $9.996839721391947880178431579096252394674 \times 10^{-15}$ \\
        $10^{15}$ & $9.998222352876870688430475630870594514676 \times 10^{-16}$ \\
        $10^{16}$ & $9.999000199970007998600279942010694075045 \times 10^{-17}$ \\
        $10^{17}$ & $9.999437721915028815978355988353941710652 \times 10^{-18}$ \\
        $10^{18}$ & $9.999683792233034558764322688103387028779 \times 10^{-19}$ \\
        $10^{19}$ & $9.999822178383382733658830066890795061768 \times 10^{-20}$ \\
        $10^{20}$ & $9.999900001999970000799986000279994220102 \times 10^{-21}$ \\
        \bottomrule
    \end{tabular}
    \end{center}
\end{table}

For various values of $M$, we numerically find the critical jumping rate by finding where $|\braket{s}{\psi_0}|^2 = |\braket{s}{\psi_1}|^2$, i.e., where the curves intersect in \fref{fig:overlaps-w=M0.75-s}. These values are shown in Table \ref{table:gamma-data-w=M0.75}. Next, we fit the function
\[
    \gamma
    =
    \frac{a_1}{M}
    +\frac{a_2}{M^{5/4}}
    +\frac{a_3}{M^{3/2}}
    +\frac{a_4}{M^{7/4}}
    +\frac{a_5}{M^{2}}
    +\frac{a_6}{M^{9/4}}
    +\frac{a_7}{M^{5/2}}
    +\frac{a_8}{M^{11/4}}
    +\frac{a_9}{M^{3}}
\]
to the data in \tref{table:gamma-data-w=M0.75} to obtain the $a_i$'s. Doing so, we numerically obtain the following critical jumping rate when $w = M^{3/4}$:
\begin{align}
    \gamma_{w=M{3/4}}
    &=
        \frac{1}{M}
        -\frac{1}{M^{5/4}}
        +\frac{2}{M^{3/2}}
        -\frac{3}{M^{7/4}}
        +\frac{8}{M^{2}} \nonumber \\
    &\quad-\frac{14}{M^{9/4}}
        +\frac{28}{M^{5/2}}
        -\frac{58}{M^{11/4}}
        +\frac{103}{M^{3}}. \label{eq:gamma-a-w=M0.75}
\end{align}
For example, when $M = 10\,000$, this yields $\gamma_{w=M{3/4}} M \approx 0.9177$, which agrees with \fref{fig:overlaps-w=M0.75}.

\begin{SCfigure}
    \caption{\label{fig:prob-time-w=M0.75}Success probability as a function of time for search on the simplex of complete graphs with $M = 10\,000$, $w = M^{3/4}$, and $\gamma = \gamma_{w=M{3/4}}$ \eqref{eq:gamma-a-w=M0.75}.}
    \includegraphics[width=2.25in]{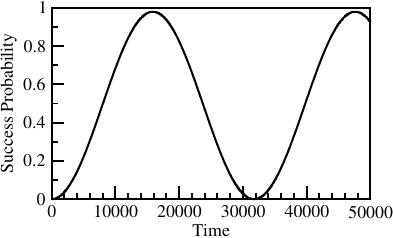}
\end{SCfigure}

\begin{table}
    \caption{\label{table:peak-success-runtime-w=M0.75}For search on the simplex of complete graphs with bridge weights $w = M^{3/4}$ and jumping rate $\gamma = \gamma_{w=M{3/4}}$, the numerically determined probability in $\ket{a}$ at time $t_{w=M{3/4}}$ \eqref{eq:t-w=M0.75}.}
    \begin{center}
    \begin{tabular}{cc}
        \toprule
        $M$ & Prob. in $\ket{a}$ at $t_{w=\sqrt{M}}$ \\
        \midrule
        $10^5$    & 0.9936330344553865847 \\
        $10^6$    & 0.9979956195096179073 \\
        $10^7$    & 0.9993671080742379587 \\
        $10^8$    & 0.9997999559499726426 \\
        $10^9$    & 0.9999367500038244599 \\
        $10^{10}$ & 0.9999799995540597370 \\
        \bottomrule
    \end{tabular}
    \end{center}
\end{table}

Using this critical jumping rate, in \fref{fig:prob-time-w=M0.75}, we plot the success probability as the system evolves with $M = 10\,000$, and it reaches a peak of nearly 1 around time $t = 15\,913.4$, which suggests that the success probability reaches 1 at time
\begin{equation}
    \label{eq:t-w=M0.75}
    t_{w=M{3/4}} = \frac{\pi}{2}\sqrt{N}.
\end{equation}
To confirm this, we can find the success probability at time $t_{w=M{3/4}}$ for various size $M$'s, and the results are shown in \tref{table:peak-success-runtime-w=M0.75}. We see that the success probability does asymptote to 1 at $t_{w=M{3/4}}$, and so the algorithm is asymptotically optimal and deterministic.


\subsection{Example: $w = M^{7/8}$}

For our third example where $\sqrt{M} \ll w \ll M$, we consider $w = M^{7/8}$.

\begin{figure}
\begin{center}
    \subfloat[] {
        \includegraphics[width=2.25in]{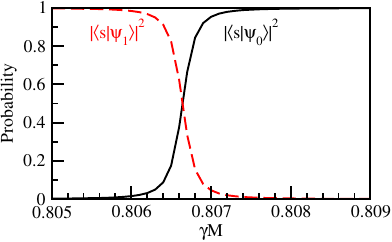}
    }
    \subfloat[] {
        \includegraphics[width=2.25in]{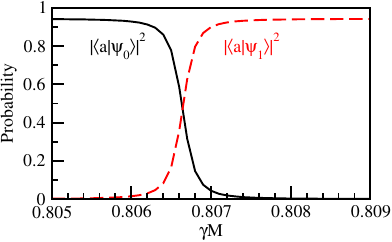}
    }

    \subfloat[] {
        \includegraphics[width=2.25in]{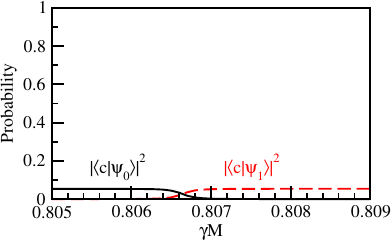}
    }
    \caption{\label{fig:overlaps-w=M0.875}For the simplex of complete graphs with $M = 10\,000$ and $w = M^{7/8}$, probability overlaps of (a) the initial state $\ket{s}$, (b) the marked vertex $\ket{a}$, and (c) $\ket{c}$ with two eigenvectors $\ket{\psi_0}$ (solid black) and $\ket{\psi_1}$ (dashed red) of the search Hamiltonian for various values of $\gamma M$.}
\end{center}
\end{figure}

We again begin by plotting the probability overlap of two eigenvectors of the search Hamiltonian \eqref{eq:H-7D} with our states of interest, which now are the initial state $\ket{s}$, the marked vertex $\ket{a}$, and the $\ket{c}$ vertex that is adjacent to $\ket{a}$ in the other clique. This is shown in \fref{fig:overlaps-w=M0.875}. We see that the critical $\gamma M$ is around 0.8066, at which the eigenvectors $\ket{\psi_0}$ and $\ket{\psi_1}$ are each half $\ket{s}$, while their other half is mostly $\ket{a}$ and a little $\ket{c}$. So, the system evolves from $\ket{s}$ to mostly $\ket{a}$ and a little $\ket{c}$.

\begin{table}
    \caption{\label{table:gamma-data-w=M0.875}For search on the simplex of complete graphs with bridge weights $w = M^{7/8}$, numerically determined values of the critical jumping rate $\gamma$ as a function of $M$.}
    \begin{center}
    \begin{tabular}{cc}
        \toprule
        $M$ & Critical $\gamma$ \\
        \midrule
        $1 \times 10^{10}$ & $9.494510249893876462571213211155143527421 \times 10^{-11}$ \\
        $2 \times 10^{10}$ & $4.766270767436967322958524632879260262832 \times 10^{-11}$ \\
        $3 \times 10^{10}$ & $3.184526928999079908341609003621225778489 \times 10^{-11}$ \\
        $4 \times 10^{10}$ & $2.391996628555953962743842836987796380096 \times 10^{-11}$ \\
        $5 \times 10^{10}$ & $1.915773844879541218495734450075280432366 \times 10^{-11}$ \\
        $6 \times 10^{10}$ & $1.597929271310387960500657286506267062430 \times 10^{-11}$ \\
        $7 \times 10^{10}$ & $1.370686949833952661454582599158714047793 \times 10^{-11}$ \\
        $8 \times 10^{10}$ & $1.200122560765524236252273741694128454037 \times 10^{-11}$ \\
        $9 \times 10^{10}$ & $1.067372560272178648149831635732318370942 \times 10^{-11}$ \\
        
        $1 \times 10^{11}$ & $9.611102815712663539556861901104391280377 \times 10^{-12}$ \\
        $2 \times 10^{11}$ & $4.820531316020564208270325839662827720563 \times 10^{-12}$ \\
        $3 \times 10^{11}$ & $3.219195527409890607840987177669241500752 \times 10^{-12}$ \\
        $4 \times 10^{11}$ & $2.417220036636699482453941147647569573737 \times 10^{-12}$ \\
        $5 \times 10^{11}$ & $1.935480153910631091570966508252840034957 \times 10^{-12}$ \\
        $6 \times 10^{11}$ & $1.614035105328532011193384940144838825728 \times 10^{-12}$ \\
        $7 \times 10^{11}$ & $1.384266208427586424079786251008192821021 \times 10^{-12}$ \\
        $8 \times 10^{11}$ & $1.211835454793997415176839318654175311166 \times 10^{-12}$ \\
        $9 \times 10^{11}$ & $1.077653012321810750498103913136088556853 \times 10^{-12}$ \\
        
        $1 \times 10^{12}$ & $9.702582564556945157915908457247582454857 \times 10^{-13}$ \\
        $2 \times 10^{12}$ & $4.862956938155323065069237177954324014674 \times 10^{-13}$ \\
        $3 \times 10^{12}$ & $3.246250506088427601868057910671515611777 \times 10^{-13}$ \\
        $4 \times 10^{12}$ & $2.436878300089401917354911986548320911777 \times 10^{-13}$ \\
        $5 \times 10^{12}$ & $1.950823388445254480629502853661757322774 \times 10^{-13}$ \\
        $6 \times 10^{12}$ & $1.626565113391620975887673555956895059959 \times 10^{-13}$ \\
        $7 \times 10^{12}$ & $1.394823658215272449457660353053774475213 \times 10^{-13}$ \\
        $8 \times 10^{12}$ & $1.220936766739738446052491280095695942367 \times 10^{-13}$ \\
        $9 \times 10^{12}$ & $1.085637378046376048007908506541956190537 \times 10^{-13}$ \\
        
        $1 \times 10^{13}$ & $9.773600200125825135665362016398378466672 \times 10^{-14}$ \\
        $2 \times 10^{13}$ & $4.895803658623360449187696550393906245527 \times 10^{-14}$ \\
        $3 \times 10^{13}$ & $3.267165750655672779054846507943365901073 \times 10^{-14}$ \\
        $4 \times 10^{13}$ & $2.452059934956175150156128683614106336835 \times 10^{-14}$ \\
        $5 \times 10^{13}$ & $1.962663529230574295472481663704511677626 \times 10^{-14}$ \\
        $6 \times 10^{13}$ & $1.636228402616895689110748906152075969851 \times 10^{-14}$ \\
        $7 \times 10^{13}$ & $1.402961546243284843486041208012771978753 \times 10^{-14}$ \\
        $8 \times 10^{13}$ & $1.227949194107014041960692280353525114782 \times 10^{-14}$ \\
        $9 \times 10^{13}$ & $1.091786895495225212885800700755833532680 \times 10^{-14}$ \\
        
        $1 \times 10^{14}$ & $9.828279403039040040034248185868603535860 \times 10^{-15}$ \\
        $2 \times 10^{14}$ & $4.921040691709277631976018521022510250510 \times 10^{-15}$ \\
        $3 \times 10^{14}$ & $3.283217024497330491133623381451131327783 \times 10^{-15}$ \\
        $4 \times 10^{14}$ & $2.463701848404867427940002196600277361265 \times 10^{-15}$ \\
        $5 \times 10^{14}$ & $1.971737681331169423334594637298231644964 \times 10^{-15}$ \\
        $6 \times 10^{14}$ & $1.643630747684189192200810779031735199415 \times 10^{-15}$ \\
        $7 \times 10^{14}$ & $1.409192954453960534289964117468448373537 \times 10^{-15}$ \\
        $8 \times 10^{14}$ & $1.233317016791335463220464232948985190719 \times 10^{-15}$ \\
        $9 \times 10^{14}$ & $1.096492820780936413108171170118072834280 \times 10^{-15}$ \\
        \bottomrule
        \multicolumn{2}{r}{{Continued on next page}} \\
    \end{tabular}
    \end{center}
\end{table}

\addtocounter{table}{-1}
\begin{table}
    \caption{Continued from previous page.}
    \begin{center}
    \begin{tabular}{cc}
        \toprule
        $M$ & Critical $\gamma$ \\
        \midrule
        $1 \times 10^{15}$ & $9.870112024759702476830158541319647756762 \times 10^{-16}$ \\
        $2 \times 10^{15}$ & $4.940317423024236667414589187822832672763 \times 10^{-16}$ \\
        $3 \times 10^{15}$ & $3.295466622116664332927272258382824494175 \times 10^{-16}$ \\
        $4 \times 10^{15}$ & $2.472581090934497963880118559990679114225 \times 10^{-16}$ \\
        $5 \times 10^{15}$ & $1.978655374920728072211662150662226421020 \times 10^{-16}$ \\
        $6 \times 10^{15}$ & $1.649271902230712538884111231439472261903 \times 10^{-16}$ \\
        $7 \times 10^{15}$ & $1.413940346234849374163800463491173307751 \times 10^{-16}$ \\
        $8 \times 10^{15}$ & $1.237405446039378187028139821441945557039 \times 10^{-16}$ \\
        $9 \times 10^{15}$ & $1.100076319722073982886554472109970126275 \times 10^{-16}$ \\
        
        $1 \times 10^{16}$ & $9.901960784313825509803921568607171175330 \times 10^{-17}$ \\
        $2 \times 10^{16}$ & $4.954975549895570777273342597344545759105 \times 10^{-17}$ \\
        $3 \times 10^{16}$ & $3.304775025846262421977823922834298344099 \times 10^{-17}$ \\
        $4 \times 10^{16}$ & $2.479325295319706455735214471357523994883 \times 10^{-17}$ \\
        $5 \times 10^{16}$ & $1.983907882863908086866193045908636913770 \times 10^{-17}$ \\
        $6 \times 10^{16}$ & $1.653553976942740669319509992439718990650 \times 10^{-17}$ \\
        $7 \times 10^{16}$ & $1.417543166358731639804242978409708268520 \times 10^{-17}$ \\
        $8 \times 10^{16}$ & $1.240507575538836167672859417407649037667 \times 10^{-17}$ \\
        $9 \times 10^{16}$ & $1.102794871888288936442463071829589329357 \times 10^{-17}$ \\
        
        $1 \times 10^{17}$ & $9.926118643101090291512662967490581353467 \times 10^{-18}$ \\
        $2 \times 10^{17}$ & $4.966083654487988410787618047897122026714 \times 10^{-18}$ \\
        $3 \times 10^{17}$ & $3.311825450623461288118430800658010085436 \times 10^{-18}$ \\
        $4 \times 10^{17}$ & $2.484431757677529753267141169956338354671 \times 10^{-18}$ \\
        $5 \times 10^{17}$ & $1.987883850292175327885775803430972965930 \times 10^{-18}$ \\
        $6 \times 10^{17}$ & $1.656794685673505261166614835000534793713 \times 10^{-18}$ \\
        $7 \times 10^{17}$ & $1.420269339874938195538124407741889473566 \times 10^{-18}$ \\
        $8 \times 10^{17}$ & $1.242854543299998510346905384780190546510 \times 10^{-18}$ \\
        $9 \times 10^{17}$ & $1.104851375705848886168543554494074349510 \times 10^{-18}$ \\
        
        $1 \times 10^{18}$ & $9.944391289005621014616235013567738669609 \times 10^{-19}$ \\
        $2 \times 10^{18}$ & $4.974479736865602254582419053735323582162 \times 10^{-19}$ \\
        $3 \times 10^{18}$ & $3.317152472134364677447972996856121115528 \times 10^{-19}$ \\
        $4 \times 10^{18}$ & $2.488288985821383870002763061911666471867 \times 10^{-19}$ \\
        $5 \times 10^{18}$ & $1.990886552840768962100240286411362242553 \times 10^{-19}$ \\
        $6 \times 10^{18}$ & $1.659241726319054817710049330665677497454 \times 10^{-19}$ \\
        $7 \times 10^{18}$ & $1.422327589738832688347523826370192903529 \times 10^{-19}$ \\
        $8 \times 10^{18}$ & $1.244626297875929511859559874297393349100 \times 10^{-19}$ \\
        $9 \times 10^{18}$ & $1.106403705967740639601089172655182950349 \times 10^{-19}$ \\
        
        $1 \times 10^{19}$ & $9.958183031048954761605049788706555146906 \times 10^{-20}$ \\
        $2 \times 10^{19}$ & $4.980813517326318784927758975053945936428 \times 10^{-20}$ \\
        $3 \times 10^{19}$ & $3.321169864415135206623041068961264679530 \times 10^{-20}$ \\
        $4 \times 10^{19}$ & $2.491197351970275058119116897822923144556 \times 10^{-20}$ \\
        $5 \times 10^{19}$ & $1.993150265972006010738844868829906310672 \times 10^{-20}$ \\
        $6 \times 10^{19}$ & $1.661086311494099375685476314417416215452 \times 10^{-20}$ \\
        $7 \times 10^{19}$ & $1.423878950971613169501158287566758042642 \times 10^{-20}$ \\
        $8 \times 10^{19}$ & $1.245961607774488951264581571605185132762 \times 10^{-20}$ \\
        $9 \times 10^{19}$ & $1.107573558369770167340321556394179821895 \times 10^{-20}$ \\
        
        $1 \times 10^{20}$ & $9.968575966436973717250841244357206748790 \times 10^{-21}$ \\
        $2 \times 10^{20}$ & $4.985584497763208124207585324699887886135 \times 10^{-21}$ \\
        $3 \times 10^{20}$ & $3.324195338914356464501728141080726622254 \times 10^{-21}$ \\
        $4 \times 10^{20}$ & $2.493387298432433278382585449927959547703 \times 10^{-21}$ \\
        $5 \times 10^{20}$ & $1.994854609248865649451291143211513050008 \times 10^{-21}$ \\
        $6 \times 10^{20}$ & $1.662474970320945625681231157314877291929 \times 10^{-21}$ \\
        $7 \times 10^{20}$ & $1.425046775567968478163497400899376474966 \times 10^{-21}$ \\
        $8 \times 10^{20}$ & $1.246966731169954191473943075562586042239 \times 10^{-21}$ \\
        $9 \times 10^{20}$ & $1.108454089295652960430785742504295957298 \times 10^{-21}$ \\
        \bottomrule
    \end{tabular}
    \end{center}
\end{table}

\begin{table}
    \caption{\label{table:peak-success-runtime-w=M0.875}For search on the simplex of complete graphs with bridge weights $w = M^{7/8}$ and jumping rate $\gamma = \gamma_{w=M{7/8}}$ \eqref{eq:gamma-w=M0.875}, numerically determined probabilities in $\ket{a}$ and $\ket{c}$ at time $t_{w = M^{7/8}}$ \eqref{eq:t-w=M0.875}.}
    \begin{center}
    \begin{tabular}{ccc}
        \toprule
        $M$ & Prob. in $\ket{a}$ at $t_{w=M{7/8}}$ & Prob. in $\ket{c}$ at $t_{w=M{7/8}}$ \\
        \midrule
        $10^{4}$  & 0.0037505949804240331 & 0.0002168253227521543 \\
        $10^{5}$  & 0.9119819223278721840 & 0.0335127262899007848 \\
        $10^{6}$  & 0.9766072953559028963 & 0.0222619094040803548 \\
        $10^{7}$  & 0.9860514488200330388 & 0.0136511904737825922 \\
        $10^{8}$  & 0.9916904833396741540 & 0.0081957910266983911 \\
        $10^{9}$  & 0.9951134607610465999 & 0.0048424398236898534 \\
        $10^{10}$ & 0.9971565903793431067 & 0.0028264608224104703 \\
        $10^{11}$ & 0.9983589888367307336 & 0.0016345939584015211 \\
        $10^{12}$ & 0.9990588694993803716 & 0.0009387484557096806 \\
        $10^{13}$ & 0.9994628334210690220 & 0.0005363022101453279 \\
        $10^{14}$ & 0.9996945128088141183 & 0.0003051807239131171 \\
        $10^{15}$ & 0.9998267452728246405 & 0.0001731483988334793 \\
        $10^{16}$ & 0.9999019438134006704 & 0.0000980199925314597 \\
        $10^{17}$ & 0.9999445908125332357 & 0.0000553970626950979 \\
        $10^{18}$ & 0.9999687268703790204 & 0.0000312691204692369 \\
        $10^{19}$ & 0.9999823652403423532 & 0.0000176334476686151 \\
        $10^{20}$ & 0.9999900626196851806 & 0.0000099369544385899 \\
        \bottomrule
    \end{tabular}
    \end{center}
\end{table}

As discussed in the introduction, \cite{Wang2020} derived a critical jumping rate $\gamma_\text{\tiny WWW}$ \eqref{eq:gamma-WWW} when $w \gg M^{3/4}$. When $M = 10\,000$ this yields $\gamma_\text{\tiny WWW} M = 0.8063$, which is too small compared to the actual value of 0.8066 from \fref{fig:overlaps-w=M0.875}. So, let us numerically determine the critical jumping rate. As in previous sections, we numerically find where $| \langle s | \psi_0 \rangle |^2 = | \langle s | \psi_1 \rangle |^2$ for various values of $M$. Whereas for previous sections, it was sufficient to use 13 data points ranging from $M = 10^8, 10^9, \dots, 10^{20}$, here we found that this was insufficient to get a precise enough critical jumping rate for large $M$. Instead, we use the 99 data points shown in \tref{table:gamma-data-w=M0.875}. Then, we fit it to the function
\begin{align*}
        &\frac{a_1}{M} + \frac{a_2}{M^{9/8}} + \frac{a_3}{M^{5/4}} + \frac{a_4}{M^{11/8}} + \frac{a_5}{M^{3/2}} 
        + \frac{a_6}{M^{13/8}} \\
        &\quad+ \frac{a_7}{M^{7/4}} + \frac{a_8}{M^{15/8}} + \frac{a_9}{M^2} + \frac{a_{10}}{M^{17/8}} + \frac{a_{11}}{M^{9/4}} + \frac{a_{12}}{M^{19/8}} \\
        &\quad+ \frac{a_{13}}{M^{5/2}} + \frac{a_{14}}{M^{21/8}} + \frac{a_{15}}{M^{11/4}} + \frac{a_{16}}{M^{23/8}} + \frac{a_{17}}{M^3}.
\end{align*}
Doing the fit and rounding each $a_i$ to the nearest integer, we get
\begin{align}
     \gamma_{w = M^{7/8}} 
        &= \frac{1}{M} - \frac{1}{M^{9/8}} + \frac{2}{M^{5/4}} - \frac{4}{M^{11/8}} + \frac{8}{M^{3/2}} - \frac{16}{M^{13/8}} \nonumber \\
        &\quad+ \frac{32}{M^{7/4}} - \frac{63}{M^{15/8}} + \frac{128}{M^2} - \frac{254}{M^{17/8}} + \frac{512}{M^{9/4}} - \frac{1079}{M^{19/8}} \nonumber \\
        &\quad+ \frac{2864}{M^{5/2}} - \frac{12075}{M^{21/8}} + \frac{60988}{M^{11/4}} - \frac{229326}{M^{23/8}} + \frac{420554}{M^3}. \label{eq:gamma-w=M0.875}
\end{align}
While this critical jumping rate is inaccurate when $M = 10\,000$, yielding $\gamma M \approx 0.8073$, it works very well for larger $M$. This is shown in \tref{table:peak-success-runtime-w=M0.875}, where at time
\begin{equation}
    \label{eq:t-w=M0.875}
    t_{w = M^{7/8}} = \frac{\pi}{2} \sqrt{N},
\end{equation}
the probability converges to $\ket{a}$. By comparison, using $\gamma_\text{\tiny WWW}$, the success probability goes to zero.


\section{\label{sec:w>>M}Search when \texorpdfstring{$w \gg M$}{w >> M}}

In this section, we consider $w \gg M$. Through trial and error, we numerically found that the critical jumping rate is
\begin{equation}
    \label{eq:gamma-w>>M}
    \gamma_{w \gg M} = \frac{M+w-2}{(M-2)(M+2w-2)},
\end{equation}
at which the quantum walk to evolves from $\ket{s}$ to half $\ket{a}$ and half $\ket{c}$ at time
\begin{equation}
    \label{eq:t-w>>M}
    t_{w \gg M} = \frac{\pi}{\sqrt{2}} \sqrt{N},
\end{equation}
for large $M$. This is reflected in \fref{fig:summary}. Next, we give three examples to support this observation.


\subsection{Example: $w = M^{5/4}$}

\begin{figure}
\begin{center}
    \subfloat[] {
        \includegraphics[width=2.25in]{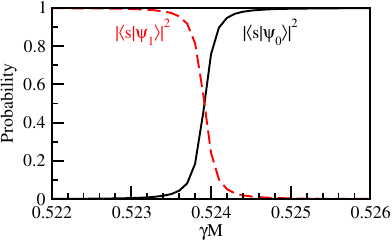}
    }
    \subfloat[] {
        \includegraphics[width=2.25in]{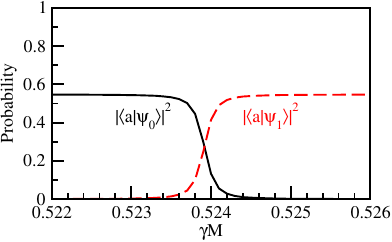}
    }

    \subfloat[] {
        \includegraphics[width=2.25in]{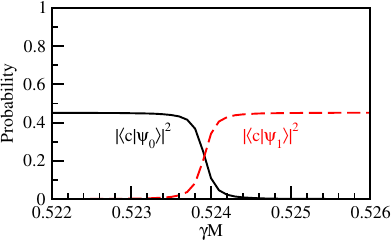}
    }
    \caption{\label{fig:overlaps-w=M1.25}For the simplex of complete graphs with $M = 10\,000$ and $w = M^{5/4}$, probability overlaps of (a) the initial state $\ket{s}$, (b) the marked vertex $\ket{a}$, and (c) $\ket{c}$ with two eigenvectors $\ket{\psi_0}$ (solid black) and $\ket{\psi_1}$ (dashed red) of the search Hamiltonian for various values of $\gamma M$.}
\end{center}
\end{figure}

\begin{SCfigure}
    \caption{\label{fig:prob-time-w=M1.25}Success probability as a function of time for search on the simplex of complete graphs with $M = 10\,000$, $w = M^{5/4}$, and $\gamma = \gamma_{w \gg M}$ \eqref{eq:gamma-w>>M}.}
    \includegraphics[width=2.25in]{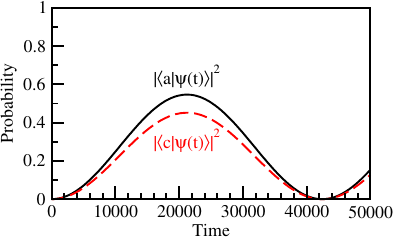}
\end{SCfigure}

\begin{table}
    \caption{\label{table:peak-success-runtime-w=M1.25}For search on the simplex of complete graphs with bridge weights $w = M^{5/4}$ and jumping rate $\gamma = \gamma_{w \gg M}$ \eqref{eq:gamma-w>>M}, numerically determined probabilities in $\ket{a}$ and $\ket{c}$ at time $t_{w \gg M}$ \eqref{eq:t-w>>M}.}
    \begin{center}
    \begin{tabular}{ccc}
        \toprule
        $M$ & Prob. in $\ket{a}$ at $t_{w \gg M}$ & Prob. in $\ket{c}$ at $t_{w \gg M}$ \\
        \midrule
        $10^5$    & 0.5261495173083902744 & 0.4716177934582705009 \\
        $10^6$    & 0.5151903194701727994 & 0.4840897283053594619 \\
        $10^7$    & 0.5086950648204609874 & 0.4910744240544123604 \\
        $10^8$    & 0.5049379159720216420 & 0.4949886444549618750 \\
        $10^9$    & 0.5027920732442048499 & 0.4971846019318747677 \\
        $10^{10}$ & 0.5015749299802275962 & 0.4984176756419047996 \\
        \bottomrule
    \end{tabular}
    \end{center}
\end{table}

The first example is when $w = M^{5/4}$, which satisfies $w \gg M$. In \fref{fig:overlaps-w=M1.25}, we plot the probability overlaps of two eigenvectors of the search Hamiltonian \eqref{eq:H-7D} with $\ket{s}$, $\ket{a}$, and $\ket{c}$ when $M = 10\,000$. We see that the critical jumping rate occurs around $\gamma M \approx 0.5239$. Our critical jumping rate \eqref{eq:gamma-w>>M} is consistent with this, as it yields $\gamma_{w \gg M} M \approx 0.5239$, whereas $\gamma_\text{\tiny WWW} M$ \eqref{eq:gamma-WWW} yields 0.5238. From \fref{fig:overlaps-w=M1.25}, at the critical jumping rate, the two eigenvectors are each half $\ket{s}$, a little over a quarter $\ket{a}$, and a little under a quarter $\ket{c}$, and so the system evolves from $\ket{s}$ to a little over half $\ket{a}$ and a little under half $\ket{c}$. This is confirmed in \fref{fig:prob-time-w=M1.25}, where at time $t_{w \gg M} \approx 22\,215$, the probability of measuring the walker at $\ket{a}$ is a bit over 50\%, while the probability of measuring the walker at $\ket{c}$ is a bit under 50\%. In \tref{table:peak-success-runtime-w=M1.25}, we give the probabilities at $\ket{a}$ and $\ket{c}$ for increasing $M$, and we see that as $M$ gets larger, the probabilities converge to 50/50.


\subsection{Example: $w = M^{3/2}$}

\begin{figure}
\begin{center}
    \subfloat[] {
        \includegraphics[width=2.25in]{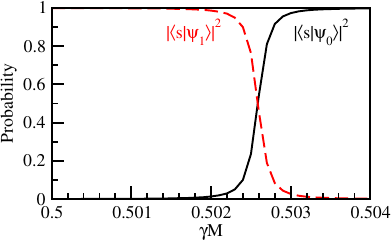}
    }
    \subfloat[] {
        \includegraphics[width=2.25in]{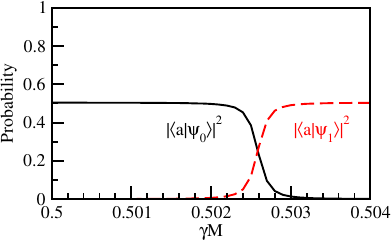}
    }

    \subfloat[] {
        \includegraphics[width=2.25in]{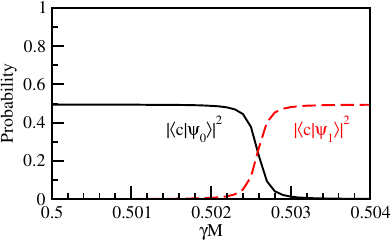}
    }
    \caption{\label{fig:overlaps-w=M1.5}For the simplex of complete graphs with $M = 10\,000$ and $w = M^{3/2}$, probability overlaps of (a) the initial state $\ket{s}$, (b) the marked vertex $\ket{a}$, and (c) $\ket{c}$ with two eigenvectors $\ket{\psi_0}$ (solid black) and $\ket{\psi_1}$ (dashed red) of the search Hamiltonian for various values of $\gamma M$.}
\end{center}
\end{figure}

\begin{SCfigure}
    \caption{\label{fig:prob-time-w=M1.5}Success probability as a function of time for search on the simplex of complete graphs with $M = 10\,000$, $w = M^{3/2}$, and $\gamma = \gamma_{w \gg M}$ \eqref{eq:gamma-w>>M}.}
    \includegraphics[width=2.25in]{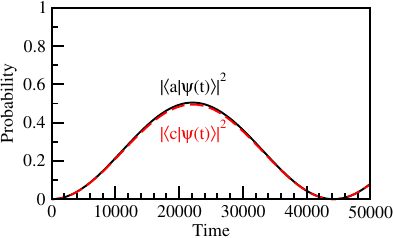}
\end{SCfigure}

\begin{table}
    \caption{\label{table:peak-success-runtime-w=M1.5}For search on the simplex of complete graphs with bridge weights $w = M^{3/2}$ and jumping rate $\gamma = \gamma_{w \gg M}$ \eqref{eq:gamma-w>>M}, numerically determined probabilities in $\ket{a}$ and $\ket{c}$ at time $t_{w \gg M}$ \eqref{eq:t-w>>M}.}
    \begin{center}
    \begin{tabular}{ccc}
        \toprule
        $M$ & Prob. in $\ket{a}$ at $t_{w \gg M}$ & Prob. in $\ket{c}$ at $t_{w \gg M}$ \\
        \midrule
        $10^5$    & 0.5015649242000018865 & 0.4984077779425919197 \\
        $10^6$    & 0.5004983781531694812 & 0.4994988837613245024 \\
        $10^7$    & 0.5001579517867516007 & 0.4998417741466419895 \\
        $10^8$    & 0.5000499837899638589 & 0.4999499887952748242 \\
        $10^9$    & 0.5000158097673819189 & 0.4999841874908860513 \\
        $10^{10}$ & 0.5000049998379067882 & 0.4999949998879119229 \\
        \bottomrule
    \end{tabular}
    \end{center}
\end{table}

For the next example of the walker asymptotically evolving to half $\ket{a}$ and half $\ket{c}$, consider $w = M^{3/2}$. With $M = 10\,000$, the probability overlaps are shown in \fref{fig:overlaps-w=M1.5}, and the critical jumping rate is around $\gamma M = 0.5026$. This agrees with $\gamma_{w \gg M} M = 0.5026$, and slightly differs from $\gamma_\text{\tiny WWW} M = 0.5025$. The runtime \eqref{eq:gamma-w>>M} is supported by \fref{fig:prob-time-w=M1.5}, where the probability in $\ket{a}$ and $\ket{c}$ each reach roughly 0.5 at time $t_{w \gg M} \approx 22\,215$. Finally, for large values of $M$, the probabilities asymptote to 0.5 each, as shown in \tref{table:peak-success-runtime-w=M1.5}.


\subsection{Example: $w = M^2$}

\begin{figure}
\begin{center}
    \subfloat[] {
        \includegraphics[width=2.25in]{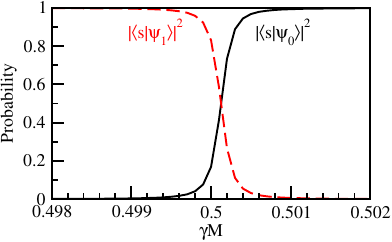}
    }
    \subfloat[] {
        \includegraphics[width=2.25in]{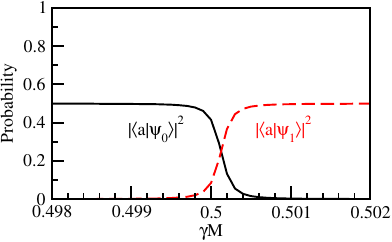}
    }

    \subfloat[] {
        \includegraphics[width=2.25in]{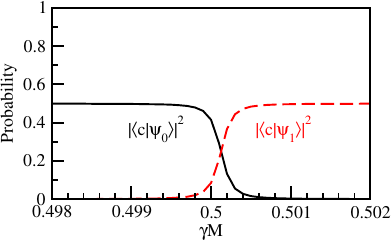}
    }
    \caption{\label{fig:overlaps-w=M2}For the simplex of complete graphs with $M = 10\,000$ and $w = M^{2}$, probability overlaps of (a) the initial state $\ket{s}$, (b) the marked vertex $\ket{a}$, and (c) $\ket{c}$ with two eigenvectors $\ket{\psi_0}$ (solid black) and $\ket{\psi_1}$ (dashed red) of the search Hamiltonian for various values of $\gamma M$.}
\end{center}
\end{figure}

\begin{SCfigure}
    \caption{\label{fig:prob-time-w=M2}Success probability as a function of time for search on the simplex of complete graphs with $M = 10\,000$, $w = M^2$, and $\gamma = \gamma_{w \gg M}$ \eqref{eq:gamma-w>>M}.}
    \includegraphics[width=2.25in]{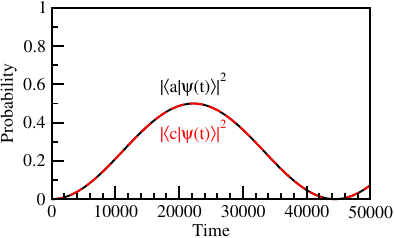}
\end{SCfigure}

\begin{table}
    \caption{\label{table:peak-success-runtime-w=M2}For search on the simplex of complete graphs with bridge weights $w = M^2$ and jumping rate $\gamma = \gamma_{w \gg M}$ \eqref{eq:gamma-w>>M}, numerically determined probabilities in $\ket{a}$ and $\ket{c}$ at time $t_{w \gg M}$ \eqref{eq:t-w>>M}.}
    \begin{center}
    \begin{tabular}{ccc}
        \toprule
        $M$ & Prob. in $\ket{a}$ at $t_{w \gg M}$ & Prob. in $\ket{c}$ at $t_{w \gg M}$ \\
        \midrule
        $10^5$    & 0.4999950003106340752 & 0.4999850008748665311 \\
        $10^6$    & 0.4999995000061141728 & 0.4999985000094489724 \\
        $10^7$    & 0.4999999500000915850 & 0.4999998500001370853 \\
        $10^8$    & 0.4999999950000005460 & 0.4999999850000011305 \\
        $10^9$    & 0.4999999995000000088 & 0.4999999985000000143 \\
        $10^{10}$ & 0.4999999999499999999 & 0.4999999998500000000 \\
        \bottomrule
    \end{tabular}
    \end{center}
\end{table}

For our last example of the algorithm evolving to equal parts $\ket{a}$ and $\ket{c}$, we consider $w = M^2$. The overlap plots are show in \fref{fig:overlaps-w=M2} with $M = 10\,000$, and the critical jumping rate corresponds to $\gamma M = 0.5001$, which is consistent with $\gamma_{w \gg M} M = 0.5001$, and slightly different from $\gamma_\text{\tiny WWW} M = 0.5000$. In \fref{fig:prob-time-w=M1.5}, probability in $\ket{a}$ and $\ket{c}$ each reach roughly 0.5 at time $t_{w \gg M} \approx 22\,215$. In \tref{table:peak-success-runtime-w=M1.5}, we see that as $M$ increases, the probabilities converge to $0.5$, and it converges more quickly than the $w = M^{5/4}$ and $w = M^{3/2}$ cases in \tref{table:peak-success-runtime-w=M1.25} and \tref{table:peak-success-runtime-w=M1.5}, respectively.


\section{\label{sec:neighbors}Inspecting Neighbors}

Recall from the introduction that when $w \ll \sqrt{M}$, it was shown in \cite{Wong7,Wong16} that the quantum walk evolves from $\ket{s}$ to $\ket{b}$ in time $O(N^{3/4}/w)$, and then a second stage of the quantum walk evolves it from $\ket{b}$ to the marked vertex $\ket{a}$ in time $O(N^{1/4})$. In this paper, we propose an alternative to this two-stage quantum walk approach. At the end of the first stage, when the quantum walk is in a uniform superposition over the $b$ vertices, we measure the position of the walker. The vertex that we find the walker at is guaranteed to be a $b$ vertex, for large $N$. From \fref{fig:simplex}, this identifies which clique contains the marked vertex. To find it, we query the oracle with each of the $M-1$ other vertices in the clique until the marked vertex is found, which will take at most $M-1 = O(\sqrt{N})$ queries. This does not affect the overall asymptotic runtime of the algorithm, which remains $O(N^{3/4}/w)$.

We note that this approach generalizes to any linear combination of $\ket{a}$ and $\ket{b}$. Measuring the position of the walker at the end of the evolution will either result in it being at the $a$ vertex or a $b$ vertex. We query the oracle with the vertex, and if it is marked, we are done. If it is not marked, it must be a $b$ vertex, and we have now identified the clique containing the marked vertex. Then, we query the oracle with each vertex in the clique until the marked vertex is found. This whole process of querying the measured position and then the other vertices in the clique takes at most $M = O(\sqrt{N})$ queries, which does not affect the overall asymptotic runtime.

Conversely, once the clique containing the marked vertex has been found, one can use a quantum walk for the second stage, following what was originally done when $w \ll \sqrt{M}$. For this second quantum-walk stage, we initialize the walker to be in a uniform superposition over the $M$ vertices of the clique containing the marked vertex. Since the initial state is roughly $\ket{b}$, it evolves to the marked vertex $\ket{a}$ in $O(N^{1/4})$ time, which does not affect the overall asymptotic runtime.

The idea of inspecting the neighbors of the measured vertex also works when the system evolves to a combination of $\ket{a}$ and $\ket{c}$. In this regime, after measuring the position of the walker, we query the oracle to determine if the vertex we found is marked or not. If it is marked, we are done. If it is unmarked, we know that we must have found vertex $c$, and so from \fref{fig:simplex}, we follow its edge to the other clique, and we must arrive at the marked vertex $a$. That is, we can infer $a$ from $c$, since the structure of the graph is known.

This seems to be the first time that inspecting the neighborhood of the walker is used with continuous-time quantum walks. The idea originates with discrete-time quantum walks searching the two-dimensional lattice \cite{Ambainis2013}.


\section{\label{sec:connectivity}Connectivity}

Intuitively, one might imagine that the higher the connectivity of a graph, the faster a quantum walk can search it. This is true of arbitrary-dimensional cubic lattices, for example, but it does not hold when comparing different families of graphs \cite{CG2004,Wong7}. For the simplex of complete graphs with weights $w \ll \sqrt{M}$, \cite{Wong16} showed that increasing the weight of the bridges both increases the connectivity and speeds up the search, which supports this intuition. In this section, however, we we will show that for even larger weights, connectivity is not a consistent indicator of fast quantum search, even within the family of weighted simplex of complete graphs.

\begin{figure}
\begin{center}
    \subfloat[] {
        \includegraphics[width=2.25in]{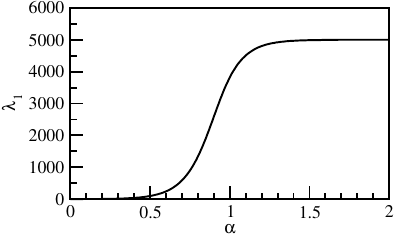}
        \label{fig:connectivity}
    }
    \subfloat[] {
        \includegraphics[width=2.25in]{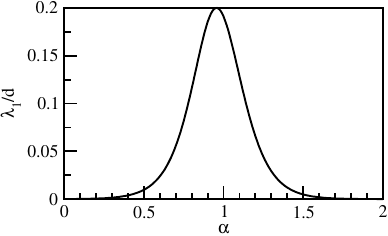}
        \label{fig:connectivity-norm}
    }
    \caption{For the simplex of complete graphs with $M = 10\,000$ and $w = M^\alpha$, the (a) algebraic connectivity and (b) normalized algebraic connectivity of the graph as $\alpha$ goes from 0 to 2.}
\end{center}
\end{figure}

A measure of connectivity that takes into account the weighted edges of the graph is the algebraic connectivity \cite{Fiedler1973}, which is the second smallest eigenvalue of the combinatorial Laplacian $L = D - A$, where $D$ is the degree matrix and $A$ is the adjacency matrix. From \cite{Wong16}, for the weighted simplex of complete graphs, the algebraic connectivity is
\[ \lambda_1 = \frac{1}{2}\left(M+2w-\sqrt{M^2-4w+4w^2}\right). \]
This is plotted in \fref{fig:connectivity} with $M = 10\,000$ and $w = M^\alpha$ as $\alpha$ goes from 0 to 2. We see that it is an increasing function that asymptotically approaches a constant. To analyze it, we take the leading-order term, which yields
\[ \lambda_1 \approx \begin{cases}
    w, & w \ll M, \\
    \frac{1 + 2c - \sqrt{1+4c^2}}{2} M, & w = cM, \\
    \frac{M}{2}, & w \gg M, \\
\end{cases} \]
where we have separated the cases where $w$ scales less than $M$, as $M$ (with a constant factor $c$), and greater than $M$. When $w \ll M$, the algebraic connectivity is approximately $w$, so it increases as $w$ gets larger. When $w \gg M$, the algebraic connectivity is approximately $M/2$, which is the value of the asymptote in \fref{fig:connectivity}.

\begin{SCfigure}
    \caption{\label{fig:runtime-alpha}For the simplex of complete graphs with $M = 10\,000$ and $w = M^\alpha$, the runtime of the search algorithm divided by $\sqrt{N}$ as $\alpha$ goes from 0 to 2.}
    \includegraphics[width=2.25in]{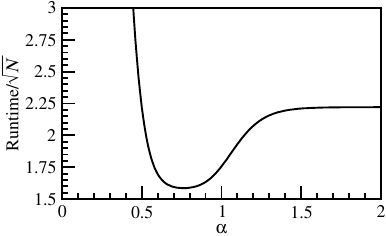}
\end{SCfigure}

To compare the algebraic connectivity with the performance of the search algorithm, we plot in \fref{fig:runtime-alpha} the runtime of the search algorithm, divided by $\sqrt{N}$, when $M = 10\,000$ and $w = M^\alpha$ with $\alpha \in [0,2]$. That is, this is a plot of the $M = 10\,000$ case from \fref{fig:heatmap-runtime}. Comparing \fref{fig:connectivity} and \fref{fig:runtime-alpha}, in the region around $\alpha = 1$, the algebraic connectivity is increasing, but the runtime is getting slower rather than speeding up. Thus, for the weighted simplex of complete graphs, higher algebraic connectivity is not a reliable indicator of faster quantum search. In other regions, however, it is a decent indicator. For $w \ll M$, the connectivity is increasing and the runtime is decreasing, and for $w \gg M$, the connectivity and the runtime are both approaching constants.

Another measure of connectivity is the normalized algebraic connectivity \cite{Chung1997}, which for a regular graph is obtained by dividing the algebraic connectivity $\lambda_1$ by the degree $d$ of the graph, which for the simplex of complete graphs is $d = M + w - 1$. This yields
\[ \frac{\lambda_1}{d} = \frac{M+2w-\sqrt{M^2-4w+4w^2}}{2(M + w - 1)}. \]
Again with $M = 10\,000$ and $w = M^\alpha$, this is plotted in \fref{fig:connectivity-norm}. The normalized algebraic connectivity reaches a maximum of
\[ \left( \frac{\lambda_1}{d} \right)_\text{max} = \frac{M+1}{5M-4} \]
when $w = (2M-1)/3$, which is when $\alpha$ equals
\[ \log_M \left( \frac{2M-1}{3} \right). \]
When $M = 10\,000$, this is a maximum of 0.20004 at $\alpha = 0.956$, in agreement with \fref{fig:connectivity-norm}. Before the maximum, the normalized algebraic connectivity is increasing, and after, it is decreasing. To get a sense of this, we examine its leading-order terms:
\[ \frac{\lambda_1}{d} \approx \begin{cases}
    \frac{w}{M}, & w \ll M, \\
    \frac{1 + 2c - \sqrt{1+4c^2}}{2(1+c)}, & w = cM, \\
    \frac{M}{2w}, & w \gg M. \\
\end{cases} \]
This shows that when $w \ll M$, the normalized algebraic connectivity decreases as $w$ increases, and when $w \gg M$, it decreases as $w$ increases.

Again, a large normalized algebraic connectivity is not a reliable indicator of fast quantum search. Comparing \fref{fig:connectivity-norm} and \fref{fig:runtime-alpha}, when $0.75 < \alpha < 0.956$, the connectivity is increasing, but the algorithm is slowing down, hence disproving the intuition. Elsewhere, the intuition holds when only considering the trends up and down. That is, when $\alpha < 0.75$, the connectivity is increasing and the algorithm is speeding up; when $\alpha > 0.956$, the connectivity is decreasing and the algorithm is slowing down; and they both asymptotically approach constants for large $\alpha$. Despite this, the actual values of the normalized algebraic connectivity are unreliable. When $\alpha = 0$ and $\alpha = 2$, the normalized algebraic connectivities are approximately equal (both zero), yet the actual runtimes are drastically different, one being $O(N^{3/4})$ and the other $O(\sqrt{N})$.

Thus, both the algebraic and normalized algebraic connectivities are not reliable indicators of fast quantum search, even within a graph family, namely the weighted simplex of complete graphs.


\section{\label{sec:conclusion}Conclusion}

In summary, we have studied the search on the weighted simplex of complete graphs for a single marked vertex. Analytically, we proved that when $w = M$, the success probability reaches 80\%, which is an improvement over previous numerical results of 36\% and 75\%. This occurs in $O(\sqrt{N})$ time, and this is the first proof of an optimal quantum search algorithm. We had several numerical results as well, which were summarized in \fref{fig:summary}. In particular, we showed that when $w \ge \sqrt{M}$, the search algorithm is optimal, although not necessarily deterministic. Furthermore, we showed that when $\sqrt{M} \ll w \ll M$, for which the search algorithm is asymptotically both optimal and deterministic. By comparison, previous results were either optimal or deterministic, not both. Next, when the search algorithm is nondeterministic, we gave a method for finding the marked vertex by inspecting vertices neighboring the measured result. Finally, the algebraic connectivity and normalized algebraic connectivity are not consistent indicators of fast quantum search, even within the family of weighted simplex of complete graphs. Analytical proofs of our numerical observations would be welcomed topics of further research.


\begin{acknowledgements}
    This material is based upon work supported in part by the National Science Foundation EPSCoR Cooperative Agreement OIA-2044049, Nebraska’s EQUATE collaboration. Any opinions, findings, and conclusions or recommendations expressed in this material are those of the author(s) and do not necessarily reflect the views of the National Science Foundation.
\end{acknowledgements}

\noindent {\small \textbf{Data Availability} \enspace No datasets were generated or analyzed during the current study.

\section*{Declarations}

\noindent {\small \textbf{Conflict of interest} \enspace K.H.F.~and Y.S.~have no competing interests to declare that are relevant to the content of this article. T.G.W.~is on the Editorial Board of the journal.}
}


\bibliographystyle{qinp}
\bibliography{refs}

\end{document}